# Integrated Heat and Power System Scheduling with Continuous-Time Thermal Dynamics via Bernstein–Galerkin Optimization

Jie Deng, Zhigang Li, *Senior Member, IEEE*, J. H. Zheng, *Member, IEEE*, and Ye Guo, *Senior Member, IEEE*

***Abstract*—Coordinated scheduling of district heating networks (DHNs) and electric power systems can improve operational flexibility and reduce costs by exploiting thermal inertia. Most existing formulations rely on simplified discrete-time DHN models, which may inadequately represent continuous spatiotemporal thermal dynamics and can lead to biased flexibility estimation and suboptimal schedules. In this paper, an integrated heat and power system scheduling framework that explicitly incorporates the continuous-time thermal dynamics of DHNs is proposed. A Bernstein–Galerkin transform method is developed to convert the underlying partial-differential thermal-dynamics constraints into a finite set of algebraic constraints, enabling tractable optimization while retaining dynamic fidelity. The resulting model transforms the original infinite-dimensional variational problem into a finite-dimensional coefficient optimization that can be solved using optimization solvers. Compared with conventional discretization approaches, the proposed method provides a more accurate representation of thermal dynamics and yields schedules with improved economic performance and reliability.**

## NOMENCLATURE

*Abbreviations*

| | |
|---|---|
| BGM | Bernstein-Galerkin method |
| CF-VT | Constant-flow variable-temperature |
| CHP | Combined heat and power |
| DHN | District heating network |
| EPS | Electric power system |
| FDM | Finite-difference method |
| IHPS | Integrated heat and power system |
| MOC | Method of characteristics |
| NM | Node method |
| ODE | Ordinary differential equation |
| PDE | Partial differential equation |
| VF-VT | Variable-flow and variable-temperature |

*Indices and Sets*

| | |
|---|---|
| $\mathcal{I}^{\mathrm{CHP}}$ | Index set of CHP units. |
| $\mathcal{I}^{\mathrm{CU}}$ | Index set of conventional non-CHP units. |
| $\mathcal{I}^{\mathrm{DP}}$ | Index set of DHN pipelines. |
| $\mathcal{I}^{\mathrm{WP}}$ | Index set of water pumps. |
| $\mathcal{I}^{\mathrm{tl}}$ | Index set of transmission lines. |
| $\mathcal{I}^{\mathrm{wind}}$ | Index set of wind farms. |
| $\mathcal{K}$ | Index set of scheduling periods. |
| $\mathcal{N}^{\mathrm{bus}}$ | Index set of buses in the EPS. |
| $\mathcal{N}^{\mathrm{line}}$ | Index set of transmission lines in the EPS. |
| $\mathcal{N}^{\mathrm{pipe}}$ | Index set of pipelines in the DHN. |
| $\mathcal{N}^{\mathrm{nd}}$ | Index set of nodes in the DHN. |
| $\mathrm{Nd}_l^{\mathrm{in}}/\mathrm{Nd}_l^{\mathrm{out}}$ | Index of the starting/ending nodes of pipeline *l*. |
| $\mathrm{Nd}_j^{\mathrm{WP}}$ | Index of the nodes connecting to water pump *j*. |
| $\mathcal{S}_i^{\mathrm{PS+}}/\mathcal{S}_i^{\mathrm{PS-}}$ | Index set of supply pipelines starting/ending at DHN node *i*. |
| $\mathcal{S}_i^{\mathrm{PR+}}/\mathcal{S}_i^{\mathrm{PR-}}$ | Index set of return pipelines starting/ending at DHN node *i*. |
| $\mathcal{S}_i^{\mathrm{HS}}/\mathcal{S}_i^{\mathrm{HL}}$ | Index set of DHN nodes associated with heat source/heat load *i*. |
| $\mathcal{S}_b^{\mathrm{CU}}/\mathcal{S}_b^{\mathrm{CHP}}$ | Index set of non-CHP/CHP units connecting to bus *b*. |
| $\mathcal{S}_b^{\mathrm{wind}}$ | Index set of wind farms connecting to bus *b*. |
| $\mathcal{S}_b^{\mathrm{WP}}$ | Index set of water pumps connecting to bus *b*. |

*Parameters and Constants*

| | |
|---|---|
| $A_l$, $\lambda_l$, $\alpha_l$ | Cross-sectional area, thermal conductivity, and thermal loss coefficient of pipeline *l*, respectively. |
| $a_{m,i}$ | Cost coefficient of CHP unit *i*. |
| $c_{m,i}$ | Cost coefficient of non-CHP unit *i*. |
| $\boldsymbol{D}_n^{\mathrm{B}}$ | Derivative matrix of the Bernstein basis. |
| $F_l$ | Transmission limit of line *l*. |
| $LD_{b,k}$ | Power load at bus *b* in period *k*. |
| $\boldsymbol{M}^{\mathrm{SP/TE}}$, $\boldsymbol{K}^{\mathrm{SP/TE}}$, $\boldsymbol{F}^{\mathrm{ST}}$ | Mass matrix, advection matrix, and load vector in the Galerkin equation, respectively. |
| $P_g^{\zeta}/Q_g^{\zeta}$ | Electricity/heat output of the $\zeta$-th extreme point in the operating region of CHP unit *g*. |
| $r_i^{\mathrm{up}}/r_i^{\mathrm{dn}}$ | Upward/downward ramping rate of unit *i*. |
| $SR^{\mathrm{up}}/SR^{\mathrm{dn}}$ | Upward/downward spinning reserve capacity requirement. |
| $SF_{l,b}$ | Shift factor of bus *b* to line *l*. |
| $T_{\mathrm{amb}}$ | Ambient temperature. |
| $T_{\mathrm{ini}}$ | Initial temperature along the pipe. |
| $\alpha_{g,k}^{\zeta}$, $NK_g$ | Output coefficients and number of extreme points for CHP unit *g*, respectively. |
| $\mu_l$ | Pressure loss coefficient of pipeline *l*. |
| $\sigma_i$ | Wind spillage penalty factor of wind farm *i*. |

This work was supported by the National Natural Science Foundation of China (52577140), Shenzhen Science and Technology Program (JCYJ20250604140301002), and Shenzhen Key Lab of Crowd Intelligence Empowered Low-Carbon Energy Network (ZDSYS20220606100601002). J. Deng and J. H. Zheng are with the School of Electric Power Engineering, South China University of Technology, Guangzhou 510641, China. Z. Li is with the School of Science and Engineering, Shenzhen Key Laboratory of Crowd Intelligence Empowered Low-Carbon Energy Network, The Chinese University of Hong Kong (Shenzhen), Shenzhen 518172, China. Y. Guo is with Department of Building Environment and Energy Engineering, The Hong Kong Polytechnic University, Hong Kong, China. (Corresponding author: Zhigang Li, lizg@cuhk.edu.cn)

*Variables*

| | |
|---|---|
| $m_l^{\mathrm{S}}, m_l^{\mathrm{R}}$ | Mass flow rate of pipeline *l* in the supply/return network. |
| $m_g^{\mathrm{G}}, m_d^{\mathrm{D}}$ | Mass flow rates of heat source *g* and heat load *d*, respectively. |
| $pr_i^{\mathrm{S}}, pr_i^{\mathrm{R}}$ | Pressure head at node *i* in the supply/return network. |
| $P_i, P_i^{\mathrm{w}}$ | Generation production of unit *i*/wind farm *i*. |
| $P_j^{\mathrm{PUMP}}$ | Power consumption of water pump *j*. |
| $Q_d$ | Heat demand of heat load *d*. |
| $Q_g$ | Heat generation of heat source *g*. |
| $ru_i, rd_i$ | Upward/downward spinning reserve of unit *i*. |
| $T_l^{\mathrm{OS}}, T_l^{\mathrm{OR}}$ | Outlet temperature of pipeline *l* in the supply/return network. |
| $T_l^{\mathrm{IS}}, T_l^{\mathrm{IR}}$ | Inlet temperature of pipeline *l* in the supply/return network. |
| $T_i^{\mathrm{NS}}, T_i^{\mathrm{NR}}$ | Temperature of node *i* in supply/return network. |
| $T_g^{\mathrm{GS}}, T_g^{\mathrm{GR}}$ | Supply/return temperature of heat source *g*. |
| $T_d^{\mathrm{DS}}, T_d^{\mathrm{DR}}$ | Supply/return temperature of heat load *d*. |
| $T_l$ | Temperature distribution of pipeline *l*. |
| $v_l$ | Flow velocity of pipeline *l*. |

## I. Introduction

The global imperative to decarbonize and mitigate climate change is driving a profound transformation in the entire energy sector from traditional separate energy systems to a deeply integrated multienergy framework [1]. In this context, integrated energy systems are gaining significant traction as key approaches, the core principle of which lies in the synergistic integration of diverse energy carriers to achieve mutual complementarity and coupling [2], [3]. This method unlocks the potential for efficiently utilizing energy by leveraging the distinct attributes of various energy vectors. Integrated heat and power system (IHPS) is a representative type of multienergy system that is distinguished by the strong coupling between the fast dynamics of an electrical power system (EPS) and the slow thermal inertia of a district heating network (DHN). The coordination of these systems has garnered significant attention from both academia and industry [4].

Leveraging the inherent flexibility of thermal systems by coordinating an IHPS is critical for achieving improved performance and economic efficiency, and many efforts have been made to harness this potential on several fronts. On the source side, the initial research concentrated on the effects of key assets such as thermal storage and heat pumps on system flexibility [5], [6]. Additionally, operational strategies for combined heat and power (CHP) units under joint dispatch were investigated to examine how the reserve capacity can be enhanced to support the IHPS [7]-[9]. On the load side, studies have focused on the management of flexible resources within the DHN [10]. With respect to operational modeling, some studies have investigated system-level unified models for optimizing the IHPS [11], [12]. These studies were predicated on steady-state or simplified models of DHNs, which neglect essential heat dynamics and may lead to unreliable or even infeasible dispatch schedules.

Moving beyond these simplified representations, subsequent studies utilized partial differential equations (PDEs) to capture the thermal dynamics of DHNs with higher fidelity. However, directly incorporating PDEs into the optimization process poses significant challenges, as the resulting complex nonlinear constraints can render the problem computationally intractable. To address this, a prevailing approach is to employ the finite difference method (FDM). This method discretizes the governing PDEs into a system of algebraic equations, effectively approximating the thermal dynamics [13]-[15]. While this method has been widely adopted, it struggles to balance modeling fidelity with computational efficiency, as inappropriate discretization schemes can introduce numerical instability. Finite-element and element-based methods have also been adopted for the dynamic thermohydraulic modeling of DHNs [16], [17]. By discretizing pipes into finite elements or segments, these methods can describe temperature propagation and heat loss in detail. However, when used in system-level scheduling, their accuracy and computational burden are governed by the selected mesh resolution, leading to a tradeoff between modeling accuracy and computational efficiency. The method of characteristics (MOC) is also widely utilized to resolve hyperbolic PDE constraints by converting PDEs into ordinary differential equations (ODEs) along characteristic lines, providing accurate solutions under constant-flow conditions [18]. Under the variable-flow operation mode, however, the characteristic lines vary with the flow trajectory, and interpolation is required to locate the upstream temperature. This procedure inevitably introduces interpolation errors and hinders direct implementation in system-level dispatch optimization models where mass flow rates are treated as decision variables. To reduce the computational burden in optimization, compact approximations such as node-based or equivalent thermal models have been commonly adopted. The node method (NM) is a popular alternative for mitigating the computational challenges associated with the FDM. Reference [19] proposed an iterative NM algorithm to solve the IHPS optimization problem, and [20] developed a quasidynamic model based on the NM to capture the thermal dynamics involved in the coordinated dispatch of distributed IHPSs. Other studies have utilized the NM to investigate the effects of thermal inertia and its potential for operational flexibility [21], [22]. As a quasidynamic paradigm, the NM simplifies the system's thermal dynamic behaviors by defining the nodal outlet temperature as a flow-weighted average over the dispatch interval. This simplification can introduce errors to the overall dispatch model. Additionally, other approaches, such as energy circuit theory [23]-[24], have been explored to model the thermal dynamics of DHNs. This theory transfers the problem to the frequency domain to solve the associated PDEs.

Although the thermal dynamics of a DHN are inherently continuous, approximating them via the aforementioned discrete-time methods introduces fundamental limitations. Discrete-time methods struggle to balance accuracy with numerical stability and scalability, leading to considerable modeling errors that may result in unreliable or infeasible dispatch schedules [25], [26]. Furthermore, inappropriate temporal resolutions limit the utilization of thermal flexibility when addressing significant load fluctuations. These limitations become more significant in the variable-flow and variable-temperature (VF-VT) operation mode. Compared with the constant-flow and variable-temperature (CF-VT) operation mode, in which the heat supply is regulated merely through controlling temperature trajectories with fixed hydraulic regimes, the VF-VT operation mode allows both mass flow rates and temperatures to affect heat supply regulation. This enlarges the feasible operation region and improves the utilization of hydraulic regulation flexibility [27]. In contrast, the constant-flow or constant-temperature assumptions, corresponding to the CF-VT and VF-CT modes, simplify the model but restrict the feasible operating region and sacrifice dispatch optimality. In the VF-VT operation mode, heat transport delays and thermal inertia are jointly affected by time-varying flow and temperature trajectories; thus, a framework that can handle spatiotemporal DHN dynamics effectively is needed.

Functional space optimization is a powerful tool for parameterizing the continuous temporal trajectories of system variables. This method models entire dynamic trajectories via a set of basis functions, thereby transforming an infinite-dimensional variational problem into a finite-dimensional coefficient optimization problem. In [28] and [29], the continuous-time trajectories of power generation and load curves in an EPS were constructed via Bernstein polynomials. This research was extended in [30] and [31] to the coordinated operations of integrated gas and power systems to increase model accuracy and dispatch flexibility. However, this approach cannot be directly exploited for DHNs because DHNs exhibit more complex dynamics with spatiotemporal coupling during heat transfer, which requires a more accurate modeling scheme to promote the exploitation of flexibility. A dedicated reformulation is therefore required to embed the PDE-governed thermal dynamic process into the IHPS scheduling framework.

To bridge these gaps, an IHPS scheduling framework considering continuous-time thermal dynamics via Bernstein–Galerkin optimization is proposed in this paper. The proposed approach models the IHPS in a continuous spatiotemporal domain to accurately capture thermal dynamics and unlock the flexibility of thermal inertia. In the proposed formulation, Bernstein basis functions parameterize both time-dependent operating trajectories and spatiotemporal states, whereas the Galerkin projection converts the PDE constraints into algebraic relations on the Bernstein coefficients. The contributions of this paper are threefold.

1) The optimization problem of IHPS is formulated in the Bernstein function space, which represents variables as continuous trajectories of spatial and temporal parameters to convert the infinite-dimensional problem to a finite-dimensional problem concerning basis coefficients. The formulation accommodates the VF-VT operation by jointly optimizing the mass flow and temperature trajectories.

2) A Bernstein–Galerkin method is developed to address the governing PDE constraints of the DHN. By leveraging the high-order continuity of the Bernstein basis functions, the proposed method represents temperature evolution in a continuous spatiotemporal domain, thereby retaining thermal dynamic fidelity and avoiding inaccuracies introduced by simplified discrete-time approximations.

3) The proposed scheduling model is validated through case studies at different scales. Compared with conventional discrete-time methods, the proposed approach characterizes thermal dynamics more accurately. Furthermore, it facilitates renewable power integration and enhances economic performance while ensuring the reliability of the heating supply.

In the remainder of this paper, Section II formulates the variational IHPS scheduling problem and analyzes its solution challenges. In Section III, the objective function and non-PDE constraints are transformed via the Bernstein polynomial basis. A Bernstein–Galerkin method is developed in Section IV for handling the PDE constraints. Case studies are analyzed in Section V. Section VI concludes the paper.

## II. Problem Formulation of Continuous Energy Flow in Integrated Heat and Power System

### *A. Variational Problem Formulation of IHPS*

The optimal operation of an IHPS is typically formulated as a nonconvex nonlinear programming problem. To capture the continuous dynamic characteristics of the IHPS, the problem is formulated as a variational problem model over a continuous time horizon, as presented below.

*1) Decision Variables*

The decision variables of an IHPS are divided into two categories on the basis of their dependencies on time and space.

The first type includes the time-dependent variables $\boldsymbol{y}(t)$. For the EPS, $\boldsymbol{y}(t)$ includes unit power outputs $P_i(t)$, wind power $P^{\mathrm{w}}(t)$, and spinning reserves $ru_i(t)$ and $rd_i(t)$. For the DHN, $\boldsymbol{y}(t)$ includes the pipeline mass flow rates $m_l^{\mathrm{S}}(t)$ and $m_l^{\mathrm{R}}(t)$, heat source and heat load mass flow rates $m_g^{\mathrm{G}}(t)$ and $m_d^{\mathrm{D}}(t)$, nodal temperatures $T_i^{\mathrm{NS}}(t)$ and $T_i^{\mathrm{NR}}(t)$, pipeline inlet and outlet temperatures $T_l^{\mathrm{IS}}(t)$, $T_l^{\mathrm{IR}}(t)$, $T_l^{\mathrm{OS}}(t)$ and $T_l^{\mathrm{OR}}(t)$, nodal pressures $pr_i^{\mathrm{S}}(t)$ and $pr_i^{\mathrm{R}}(t)$, water pump power consumption $P_j^{\mathrm{PUMP}}(t)$, thermal outputs of the CHP units $Q_i(t)$, heat source outputs $Q_g(t)$, thermal exchange of the load $Q_d(t)$, and pipeline flow velocity $v_l(t)$.

The other type refers to the spatiotemporal variables $\boldsymbol{z}(x, t)$, including the pipeline temperature $T_l(x, t)$, whose evolution describes the heat transfer process across the network.

*2) Objective Function*

The objective function $J$ seeks to minimize the total operation cost, including the operation costs of non-CHP units $C_i^{\mathrm{CU}}$, CHP units $C_i^{\mathrm{CHP}}$, and wind power curtailment penalties $C_i^{\mathrm{wind}}$:

$$\min J = \sum_{i\in\mathcal{I}^{\mathrm{CU}}} C_i^{\mathrm{CU}} + \sum_{i\in\mathcal{I}^{\mathrm{wind}}} C_i^{\mathrm{wind}} + \sum_{i\in\mathcal{I}^{\mathrm{CHP}}} C_i^{\mathrm{CHP}}, \tag{1}$$

$$C_i^{\mathrm{CU}} = c_{0,i} + c_{1,i}\int_0^{\mathcal{T}} P_i(t)\,dt + c_{2,i}\left[\int_0^{\mathcal{T}} P_i(t)\,dt\right]^2, \forall i \in \mathcal{I}^{\mathrm{CU}}, \tag{2}$$

$$C_i^{\text{wind}} = \int_0^{\mathcal{T}} \sigma_i \left[ \bar{P}_i^{\text{w}}(t) - P_i^{\text{w}}(t) \right] dt, \ \forall i \in \mathcal{I}^{\text{wind}}, \tag{3}$$

$$\begin{aligned} C_i^{\text{CHP}} &= a_{0,i} + a_{1,i} \int_0^{\mathcal{T}} P_i(t)\, dt + a_{2,i} \int_0^{\mathcal{T}} Q_i(t)\, dt \\ &+ a_{3,i} \left[ \int_0^{\mathcal{T}} P_i(t)\, dt \right]^2 + a_{4,i} \left[ \int_0^{\mathcal{T}} Q_i(t)\, dt \right]^2, \ \forall i \in \mathcal{I}^{\text{CHP}}, \end{aligned} \tag{4}$$

where $C_i^{\text{CU}}$ and $C_i^{\text{CHP}}$ are defined as quadratic functions of their power outputs and $\mathcal{T}$ is the continuous scheduling time horizon.

*3) EPS Constraints*

The following constraints define the EPS feasible region.

$$\begin{aligned} &\sum\nolimits_{i \in \mathcal{I}^{\text{CU}} \cup \mathcal{I}^{\text{CHP}}} P_i(t) + \sum\nolimits_{i \in \mathcal{I}^{\text{wind}}} P_i^{\text{w}}(t) - \sum\nolimits_{j \in \mathcal{I}^{\text{WP}}} P_j^{\text{PUMP}}(t) \\ &= \sum\nolimits_{b \in \mathcal{N}^{\text{bus}}} LD_b(t), \ \forall t \in \mathcal{T}, \end{aligned} \tag{5}$$

$$\begin{aligned} &\Big| \sum\nolimits_{b \in \mathcal{N}^{\text{bus}}} SF_{l,b} \Big[ \sum\nolimits_{i \in \mathcal{S}_b^{\text{CU}} \cup \mathcal{S}_b^{\text{CHP}}} P_i(t) + \sum\nolimits_{i \in \mathcal{S}_b^{\text{wind}}} P_i^{\text{w}}(t) - \\ &\sum\nolimits_{j \in \mathcal{S}_b^{\text{WP}}} P_j^{\text{PUMP}}(t) - \sum\nolimits_{b \in \mathcal{N}^{\text{bus}}} LD_b(t) \Big] \Big| \le F_l, \ \forall l \in \mathcal{I}^{\text{tl}}, t \in \mathcal{T}, \end{aligned} \tag{6}$$

$$\underline{P}_i \le P_i(t) \le \bar{P}_i, \ \forall i \in \mathcal{I}^{\text{CU}} \cup \mathcal{I}^{\text{CHP}}, t \in \mathcal{T}, \tag{7}$$

$$0 \le P_i^{\text{w}}(t) \le \bar{P}_i^{\text{w}}, \ \forall i \in \mathcal{I}^{\text{wind}}, t \in \mathcal{T}, \tag{8}$$

$$\begin{aligned} &0 \le ru_i(t) \le r_i^{\text{up}}, \ \forall i \in \mathcal{I}^{\text{CU}}, t \in \mathcal{T}, \\ &ru_i(t) \le \bar{P}_i - P_i(t), \ \forall i \in \mathcal{I}^{\text{CU}}, t \in \mathcal{T}, \end{aligned} \tag{9}$$

$$\begin{aligned} &0 \le rd_i(t) \le r_i^{\text{dn}}, \ \forall i \in \mathcal{I}^{\text{CU}}, t \in \mathcal{T}, \\ &rd_i(t) \le P_i(t) - \underline{P}_i, \ \forall i \in \mathcal{I}^{\text{CU}}, t \in \mathcal{T}, \end{aligned} \tag{10}$$

$$\begin{aligned} &\sum\nolimits_{i \in \mathcal{I}^{\text{CU}}} ru_i(t) \ge SR^{\text{up}}, \ \forall t \in \mathcal{T}, \\ &\sum\nolimits_{i \in \mathcal{I}^{\text{CU}}} rd_i(t) \ge SR^{\text{dn}}, \ \forall t \in \mathcal{T}, \end{aligned} \tag{11}$$

$$-r_i^{\text{dn}} \le dP_i(t)/dt \le r_i^{\text{up}}, \ \forall i \in \mathcal{I}^{\text{CU}} \cup \mathcal{I}^{\text{CHP}}, t \in \mathcal{T}. \tag{12}$$

Constraint (5) enforces the power balance, while (6) ensures transmission security by restricting line flows. Constraints (7)-(8) define the capacity limits of the generation units. Constraints (9)-(11) specify the spinning reserve requirements, and (12) dictates the ramping rate limits of the units.

*4) DHN Constraints*

The following equations model the physical constraints governing the thermal–hydraulic dynamics of the DHN.

$$\begin{aligned} &\sum\nolimits_{l \in \mathcal{S}_i^{\text{PS}-}} T_l^{\text{OS}}(t) \cdot m_l^{\text{S}}(t) + \sum\nolimits_{g \in \mathcal{S}_i^{\text{HS}}} T_g^{\text{GS}}(t) \cdot m_g^{\text{G}}(t) \\ &= T_i^{\text{NS}}(t) \Big[ \sum\nolimits_{l \in \mathcal{S}_i^{\text{PS}-}} m_l^{\text{S}}(t) + \sum\nolimits_{g \in \mathcal{S}_i^{\text{HS}}} m_g^{\text{G}}(t) \Big], \\ &\sum\nolimits_{l \in \mathcal{S}_i^{\text{PR}-}} T_l^{\text{OR}}(t) \cdot m_l^{\text{R}}(t) + \sum\nolimits_{d \in \mathcal{S}_i^{\text{HL}}} T_d^{\text{DR}}(t) \cdot m_d^{\text{D}}(t) \\ &= T_i^{\text{NR}}(t) \Big[ \sum\nolimits_{l \in \mathcal{S}_i^{\text{PR}-}} m_l^{\text{R}}(t) + \sum\nolimits_{d \in \mathcal{S}_i^{\text{HL}}} m_d^{\text{D}}(t) \Big], \\ &\forall i \in \mathcal{N}^{\text{nd}}, t \in \mathcal{T}, \end{aligned} \tag{13}$$

$$\begin{aligned} &T_l^{\text{IS}}(t) = T_i^{\text{NS}}(t), \ \forall i \in \mathcal{N}^{\text{nd}}, l \in \mathcal{S}_i^{\text{PS}+}, t \in \mathcal{T}, \\ &T_l^{\text{IR}}(t) = T_i^{\text{NR}}(t), \ \forall i \in \mathcal{N}^{\text{nd}}, l \in \mathcal{S}_i^{\text{PR}+}, t \in \mathcal{T}, \end{aligned} \tag{14}$$

$$\begin{aligned} &\sum_{l \in \mathcal{S}_i^{\text{PS}-}} m_l^{\text{S}}(t) + \sum_{g \in \mathcal{S}_i^{\text{HS}}} m_g^{\text{G}}(t) = \sum_{l \in \mathcal{S}_i^{\text{PS}+}} m_l^{\text{S}}(t) + \sum_{d \in \mathcal{S}_i^{\text{HL}}} m_d^{\text{D}}(t), \\ &\sum_{l \in \mathcal{S}_i^{\text{PR}-}} m_l^{\text{R}}(t) + \sum_{d \in \mathcal{S}_i^{\text{HL}}} m_d^{\text{D}}(t) = \sum_{l \in \mathcal{S}_i^{\text{PS}+}} m_l^{\text{R}}(t) + \sum_{g \in \mathcal{S}_i^{\text{HS}}} m_g^{\text{G}}(t), \\ &\forall i \in \mathcal{N}^{\text{nd}}, t \in \mathcal{T}, \end{aligned} \tag{15}$$

$$\begin{aligned} &pr_{i1}^{\text{S}}(t) - pr_{i2}^{\text{S}}(t) = \mu_l \cdot \left[ m_l^{\text{S}}(t) \right]^2, \\ &pr_{i1}^{\text{R}}(t) - pr_{i2}^{\text{R}}(t) = \mu_l \cdot \left[ m_l^{\text{R}}(t) \right]^2, \\ &\forall i1 = \text{Nd}_l^{\text{in}}, i2 = \text{Nd}_l^{\text{out}}, l = (i1, i2) \in \mathcal{N}^{\text{pipe}}, t \in \mathcal{T}, \end{aligned} \tag{16}$$

$$\begin{aligned} &P_j^{\text{PUMP}}(t) = 1/\rho\eta_j \cdot m_j^{\text{G}}(t) \left[ pr_i^{\text{S}}(t) - pr_i^{\text{R}}(t) \right], \\ &\forall j \in \mathcal{I}^{\text{WP}}, i \in \text{Nd}_j^{\text{WP}}, t \in \mathcal{T}, \end{aligned} \tag{17}$$

$$\begin{aligned} &P_g(t) = \sum_{\zeta=1}^{NK_g} \alpha_g^{\zeta}(t) P_g^{\zeta}(t), \ Q_g(t) = \sum_{\zeta=1}^{NK_g} \alpha_g^{\zeta}(t) Q_g^{\zeta}(t), \forall t \in \mathcal{T} \\ &\forall g \in \mathcal{I}^{\text{CHP}}, \ \sum\nolimits_{\zeta=1}^{NK_g} \alpha_g^{\zeta} = 1, \ \alpha_g^{\zeta} \in [0, \ 1], \ \zeta \in \{1, \cdots NK_g\}, \end{aligned} \tag{18}$$

$$Q_g(t) = c \cdot m_g^{\text{G}}(t) \cdot \left[ T_g^{\text{GS}}(t) - T_g^{\text{GR}}(t) \right], \ \forall g \in \mathcal{S}^{\text{HS}}, t \in \mathcal{T}, \tag{19}$$

$$Q_d(t) = c \cdot m_d^{\text{D}}(t) \cdot \left[ T_d^{\text{DS}}(t) - T_d^{\text{DR}}(t) \right], \ \forall d \in \mathcal{S}^{\text{HL}}, t \in \mathcal{T}, \tag{20}$$

$$\begin{aligned} &\frac{\partial T_l(x,t)}{\partial t} + v_l(t) \frac{\partial T_l(x,t)}{\partial x} = \alpha_l \left[ T_{\text{amb}} - T_l(x,t) \right], \\ &\alpha_l = \frac{\lambda_l}{\rho c A_l}, \ \forall l \in \mathcal{I}^{\text{DP}}, t \in \mathcal{T}. \end{aligned} \tag{21}$$

Constraint (13) defines the temperature mixing at each node, constraint (14) defines the pipeline inlet temperature as the node mixing temperature, and (15) specifies the mass flow balance. The pressure drop along each pipeline is modeled in (16). The power consumption of the heat pumps is formulated in (17), and the operational model of the CHP units is specified in (18). The heat power exchanges at heat sources and loads are represented in (19) and (20), respectively. The crucial spatiotemporal evolution of the water temperature along the pipelines is governed by the first-order PDE shown in (21).

## *B. Solution Challenges*

The variational problem formulated in (1)-(21) provides a precise and comprehensive representation of the system dynamics. However, it introduces formidable theoretical and computational challenges that render a direct numerical solution intractable, primarily stemming from two issues.

*1) Infinite-dimensional problem structure*: The decision variables are defined on the continuous domain rather than the finite-dimensional vectors in $\mathbb{R}^n$. Consequently, the constraints must be satisfied over a continuum (e.g., for all $t \in \mathcal{T}$), resulting in an uncountably infinite set of conditions, making the conventional algebraic algorithms computationally intractable.

*2) PDE-governed thermal dynamics*: The governing PDEs of thermal dynamics extend the problem from a temporal domain to a spatiotemporal domain. For a DHN operating under variable-flow conditions, the advective velocity in (21) becomes time dependent, so the heat-transport process cannot be adequately characterized by temporal variables alone. Consequently, discretizing the PDE-governed spatiotemporal states via fixed grids renders the model highly sensitive to the chosen resolution. A coarse resolution may distort thermal transport, whereas a refined resolution increases the number of algebraic variables and constraints. Crucially, PDEs are intrinsically embedded within a variational problem framework, and their state acts as a direct variable within the algebraic constraints that must hold at every instant t. This structure precludes the direct application of standard algorithms designed for finite-dimensional programming.

These challenges necessitate a method capable of preserving the continuous spatiotemporal description of thermal dynamics in a DHN while converting the infinite-dimensional variational problem into a numerically tractable algebraic problem.

## III. Bernstein-Based Transformation of the Continuous Spatiotemporal Model

The infinite dimensionality of the variational problem in Section II precludes a direct numerical solution. While conventional time-stepping discretization methods can provide finite-dimensional approximations, the piecewise-constant representation fails to preserve the inherent continuity of the system states. Instead of defining time-stepping variables at only selected instants, the Bernstein polynomials define continuous trajectories over each interval, which provides a compact representation for the subsequent coefficient optimization.

This section describes the Bernstein-based transformation methodology. Section III-A presents the Bernstein function space and formalizes the parameterization of trajectories via Bernstein coefficients. On this basis, the original infinite-dimensional problem is transformed into an algebraic optimization model in Section III-B.

### *A. Properties of Bernstein Function Space*

The time and space parameters are normalized within [0, 1] to fit in the support of the Bernstein polynomials. For each time interval $t \in [0, \Delta t]$ and length $x \in [0, L]$, the normalized coordinates are denoted as $\tau = t/\Delta t$ and $\xi = x/L$, respectively.

For a given degree $n \in \mathbb{N}_0 = \{0, 1, 2, \cdots\}$, there exists a set of $n$+1 Bernstein basis polynomials, i.e., $\{B_{i,n}(\tau)\}_{i=0}^{n}$. Each basis polynomial over $\tau \in [0, 1]$ is defined as follows:

$$B_{i,n}(\tau) = C_n^i \tau^n (1-\tau)^{n-i}, \ \forall i \in \{0, \cdots, n\}, \ n \in \mathbb{N}_0. \tag{22}$$

The function space constructed from the Bernstein basis polynomials is referred to as the Bernstein function space, which is linearly spanned by the $n$+1 Bernstein basis functions:

$$\mathcal{B}^n = \operatorname{span}\{B_{0,n}(\tau), B_{1,n}(\tau), \cdots, B_{n,n}(\tau)\}, \ \forall n \in \mathbb{N}_0. \tag{23}$$

Within this Bernstein function space, any time-dependent trajectory $\boldsymbol{y}(\tau)$ can be parameterized as a Bézier curve:

$$\boldsymbol{y}(\tau) = \sum_{i=0}^{n} y_i^{\mathrm{B}} B_{i,n}(\tau) = \left(\boldsymbol{y}^{\mathrm{B}}\right)^{\mathrm{T}} \boldsymbol{B}_n(\tau), \ \forall n \in \mathbb{N}_0, \tag{24}$$

where $\boldsymbol{y}^{\mathrm{B}} = [y_0^{\mathrm{B}}, y_1^{\mathrm{B}}, \cdots, y_n^{\mathrm{B}}]^{\mathrm{T}} \in \mathbb{R}^{(n+1)}$ is a vector of Bernstein coefficients. $\boldsymbol{B}_n(\tau) = [B_{0,n}(\tau), B_{1,n}(\tau), \cdots, B_{n,n}(\tau)]^{\mathrm{T}}$ denotes a vector of the Bernstein basis polynomials.

Similarly, the Bernstein function basis can be extended to spatiotemporal domains, forming Bézier surfaces. A spatiotemporal variable $\boldsymbol{z}(\xi, \tau)$ can be formulated as follows:

$$\begin{aligned} \boldsymbol{z}(\xi,\tau) &= \sum_{j=0}^{q}\sum_{i=0}^{n} B_{j,q}(\xi) z_{j,i}^{\mathrm{B}} B_{i,n}(\tau) \\ &= \boldsymbol{B}_q(\xi)^{\mathrm{T}} \boldsymbol{z}^{\mathrm{B}} \boldsymbol{B}_n(\tau), \ \forall q, n \in \mathbb{N}_0, \end{aligned} \tag{25}$$

where $\boldsymbol{z}^{\mathrm{B}} \in \mathbb{R}^{(q+1)\times(n+1)}$ represents the coefficient matrix in the Bernstein function space. $\boldsymbol{B}_q(\xi)$ is the basis vector in the normalized spatial domain $\xi$. For notational clarity, we denote the coefficient vector for any decision trajectory $\Gamma$ as $\boldsymbol{\Gamma}^{\mathrm{B}}$.

The use of the Bernstein basis in optimization is rooted in its fundamental mathematical properties.

*1) Convex Hull Property*: A Bézier curve or surface is contained within the convex hull of its basis coefficients [32]. An upper-bound constraint on the trajectory $y(\tau)$ can be satisfied by bounding the maximum coefficient:

$$\begin{aligned} &\boldsymbol{y}(\tau) \le C \Leftrightarrow \left(\boldsymbol{y}^{\mathrm{B}}\right)^{\mathrm{T}} \boldsymbol{B}_n(\tau) \le C \Leftarrow y_i^{\mathrm{B}} \le C, \\ &\forall i \in \{0, \cdots, n\}, \ n \in \mathbb{N}_0. \end{aligned} \tag{26}$$

*2) Calculus Properties*: Bernstein polynomials exhibit structural closure under differentiation and integration, which facilitates the handling of dynamic constraints and integral objectives. The key transformations are summarized below:

$$\begin{aligned} d\boldsymbol{y}(\tau)/d\tau &= n\sum\nolimits_{i=0}^{n-1}\left(y_{i+1}^{\mathrm{B}} - y_i^{\mathrm{B}}\right) B_{i,n-1}(\tau) \\ &= \left(\boldsymbol{D}_n^{\mathrm{B}} \boldsymbol{y}^{\mathrm{B}}\right)^{\mathrm{T}} \boldsymbol{B}_{n-1}(\tau), \ \forall n \in \mathbb{N}_0, \ n \ge 1, \end{aligned} \tag{27}$$

$$\begin{aligned} &\int_0^1 \boldsymbol{y}(\tau)\,\mathrm{d}\tau = \left(\boldsymbol{y}^{\mathrm{B}}\right)^{\mathrm{T}} \int_0^1 \boldsymbol{B}_n(\tau)\,\mathrm{d}\tau = \mathbf{1}^{\mathrm{T}} \boldsymbol{y}^{\mathbf{B}} / (n+1), \\ &\mathbf{1} = [1 \ \ 1 \ \ \cdots \ \ 1]_{(n+1)\times 1}^{\mathrm{T}}, \ \forall n \in \mathbb{N}_0, \end{aligned} \tag{28}$$

### *B. Transformation of the Variational Problem*

The ordinary differential constraints or integral objective function of the IHPS model described above can be transformed into algebraic forms by applying the Bernstein basis properties described in Section III-A.

*1) Transformation of Variables and Objectives*

Each continuous decision trajectory is uniquely represented by a finite set of Bernstein coefficients. This defines a mapping from the infinite-dimensional space of functions to a finite-dimensional Bernstein function space.

$$\boldsymbol{y}(\tau) \to \boldsymbol{y}^{\mathrm{B}} \in \mathbb{R}^{n+1}, \boldsymbol{z}(\xi,\tau) \to \boldsymbol{z}^{\mathrm{B}} \in \mathbb{R}^{(q+1)\times(n+1)}, \forall n, q \in \mathbb{N}_0. \tag{29}$$

The objective function is transformed into an algebraic function of these coefficients by applying the integration property (28).

$$\begin{aligned} &J = \alpha + \int_0^1 \beta \boldsymbol{y}(\tau)\,d\tau + \chi\left[\int_0^1 \boldsymbol{y}(\tau)\,d\tau\right]^2, \\ &\Leftrightarrow J = \alpha + \frac{\beta}{n+1}\cdot \mathbf{1}^{\mathrm{T}} \boldsymbol{y}^{\mathrm{B}} + \frac{\chi}{(n+1)^2}\left(\mathbf{1}^{\mathrm{T}} \boldsymbol{y}^{\mathrm{B}}\right)^2, \ \forall n \in \mathbb{N}_0, \end{aligned} \tag{30}$$

*2) Transformation of Non-PDE Constraints*

**Equality Constraints:** The integration property (28) is applied to convert the functional equality constraints into systems of algebraic equations. The transformation of the linear balance equations (5) and (15) over a normalized domain is formulated as follows:

$$\int_0^1 \boldsymbol{L}\boldsymbol{y}(\tau)\,d\tau = \boldsymbol{e} \Leftrightarrow \frac{1}{n+1}\boldsymbol{L}\cdot \mathbf{1}^{\mathrm{T}} \boldsymbol{y}^{\mathrm{B}} = \boldsymbol{e}, \ \forall n \in \mathbb{N}_0. \tag{31}$$

**Inequality Constraints:** The convex hull property (26) is applied to convert the inequality constraints (6)-(11) into linear inequalities over the Bernstein function space coefficients.

$$\begin{aligned} &\boldsymbol{A}\boldsymbol{y}(\tau) \le \boldsymbol{b} \Leftarrow \boldsymbol{A}\boldsymbol{y}^{\mathrm{B}} \le \boldsymbol{b} \Leftrightarrow \boldsymbol{A}y_i^{\mathrm{B}} \le \boldsymbol{b}, \\ &\forall i \in \{0, \cdots, n\}, \ n \in \mathbb{N}_0. \end{aligned} \tag{32}$$

**Differential Constraints:** The differentiation property (27) converts ordinary differential constraints, such as the ramping limits (12), into equivalent linear algebraic forms.

$$\boldsymbol{C}\frac{d\boldsymbol{y}(\tau)}{d\tau} \le \boldsymbol{d} \Leftarrow \boldsymbol{C}\boldsymbol{D}_n^{\mathrm{B}} \boldsymbol{y}^{\mathrm{B}} \le \boldsymbol{d}, \ \forall n \in \mathbb{N}_0. \tag{33}$$

For nonlinear constraints involving product terms, such as the quadratic pressure drop in (16) and the bilinear heat transfers in (13) and (19)-(20), the transformation implemented over the standard interval results in quadratic or bilinear algebraic equalities.

$$\begin{aligned}&\int_0^1 \boldsymbol{y}_1(\tau)\,d\tau\cdot\boldsymbol{W}\cdot\int_0^1 \boldsymbol{y}_2(\tau)\,d\tau=\boldsymbol{h}\\&\Leftrightarrow 1/(n+1)^2\cdot\left(\mathbf{1}^{\mathrm{T}}\boldsymbol{y}_1^{\mathrm{B}}\right)^{\mathrm{T}}\boldsymbol{W}\left(\mathbf{1}^{\mathrm{T}}\boldsymbol{y}_2^{\mathrm{B}}\right)^{\mathrm{T}}=\boldsymbol{h},\ \forall n\in\mathbb{N}_0.\end{aligned}\tag{34}$$

To this point, all algebraic and ordinary differential constraints except for (21) have been in standard algebraic form. The PDE constraint (21) governing the core thermal dynamics requires a dedicated approach and is detailed in the next section.

## IV. Bernstein–Galerkin Method for the Reformulation of Thermal Dynamics Constraints

Analytical solutions to the PDEs of DHNs are seldom possible for problems involving time-varying advective transport and other boundary conditions. Consequently, a Bernstein–Galerkin method via the Bernstein basis is proposed to obtain an approximate time-varying solution in this section.

Section IV-A outlines the fundamentals of the Galerkin method. In Section IV-B, the proposed Bernstein–Galerkin formulation is presented to project the PDE constraints onto the Bernstein function basis, yielding a system of ODEs for the time-varying Bernstein coefficients. Section IV-C summarizes the resulting finite-dimensional formulation of the continuous spatiotemporal optimization model constructed in the Bernstein function space.

### A. Fundamentals of the Galerkin Method

The Galerkin method is a class of numerical technique that is used to convert a continuous operator equation defined in an infinite-dimensional function space into a finite-dimensional algebraic system [33], [34]. Consider a general form of a PDE in a Hilbert space $\mathcal{H}$:

$$\mathcal{D}(u)=f_{\mathrm{sc}},\ \forall u\in\mathcal{H}\ ,\tag{35}$$

where $u$ is an unknown function in $\mathcal{H}$, $\mathcal{D}$ is a differential equation involving derivatives with respect to time and space, and $f_{\mathrm{sc}}$ is a known source term.

The Galerkin method seeks to obtain an approximate solution $u^*$ within a finite-dimensional subspace $\mathcal{V}_{\mathcal{H}}\subset\mathcal{H}$, which is spanned by prechosen basis functions. An approximate solution is represented as a linear combination of these basis functions:

$$u^*=\sum\nolimits_{i=0}^{N}u_i\cdot\psi_i,\ \boldsymbol{\psi}=\left[\psi_0,\psi_1,\cdots,\psi_N\right]^{\mathrm{T}},\tag{36}$$

where $u_i$ denotes the unknown coefficients and $\{\psi_i\}_{i=0}^{N}$ is a set of basis functions chosen to span the subspace $\mathcal{V}_{\mathcal{H}}$.

The Galerkin method enforces the residual $R^*=\mathcal{D}u^*-f_{\mathrm{sc}}$ to be orthogonal to the entire subspace $\mathcal{V}_{\mathcal{H}}$, minimizing the approximation error. This condition is satisfied when the residual is orthogonal to each chosen basis function:

$$\left\langle \mathcal{D}(u^*)-f_{\mathrm{sc}},\ \psi_i\right\rangle_{\mathcal{H}}=0,\ \forall i\in\mathbb{N}_0,\tag{37}$$

This orthogonality property facilitates the transformation of the infinite-dimensional PDE into a finite-dimensional algebraic system and guarantees that the solution is the best possible approximation within the chosen subspace.

### B. Galerkin Projection of PDE-Governed Thermal dynamics

Although the Galerkin method provides a general framework for transforming PDEs into finite-dimensional systems, the basis functions determine whether the resulting formulation is compatible with the scheduling model. With respect to the thermal dynamics of the DHN shown in (21), the pipe-temperature field depends on both the normalized spatial and temporal coordinates, and the advective term is coupled with the time-varying flow velocity in the VF-VT operation mode. To address this issue, a Bernstein–Galerkin method (BGM) is developed by using Bernstein basis polynomials in the Galerkin projection. This scheme reformulates the PDE residual into algebraic functions of Bernstein coefficients, yielding a continuous spatiotemporal representation consistent with the Bernstein transform in Section III.

The scheduling horizon $\mathcal{T}$ is discretized into $N_k$ uniform intervals, which are indexed by $k\in\mathcal{K}=\{1, 2, \dots, N_k\}$. Within each interval $k$, the infinite-dimensional temperature field $T(x, t)$ is approximated in the Bernstein function space:

$$T_k^*(\xi,\tau)=\boldsymbol{B}_q(\xi)^{\mathrm{T}}\boldsymbol{T}_k^{\mathrm{B}}\boldsymbol{B}_n(\tau),\ \forall q,n\in\mathbb{N}_0,k\in\mathcal{K},\tag{38}$$

where $\boldsymbol{T}_k^{\mathrm{B}}\in\mathbb{R}^{(q+1)\times(n+1)}$ is the coefficient matrix that defines the spatiotemporal surface of the temperature.

Substituting (38) into (21) yields the residual $R_k(\xi, \tau)$. By enforcing this residual to be orthogonal to the spatiotemporal basis, the orthogonality condition can be formulated as follows:

$$\int_0^1\!\!\int_0^1 R_k(\xi,\tau)B_{w,q}(\xi)\,B_{u,n}(\tau)\,d\xi d\tau=0,\ \forall w,u\in\mathbb{N}_0,\tag{39}$$

where $B_{w,q}(\xi)B_{u,n}(\tau)$ denotes the weighting functions in the Bernstein function space.

This projection transforms the differential operator terms in (21) into algebraic matrix operations as follows:

$$\begin{aligned}&\frac{1}{\Delta t}\sum_{j=0}^{q}\sum_{i=0}^{n}\left(\boldsymbol{T}_k^{\mathrm{B}}\right)_{j,i}\left[\int_0^1 B_{j,q}(\xi)\,B_{w,q}(\xi)\,d\xi\right]\\&\times\left[\int_0^1\left[dB_{i,n}(\tau)/d\tau\right]B_{u,n}(\tau)\,d\tau\right]=\frac{1}{\Delta t}\left(\boldsymbol{M}^{\mathrm{SP}}\boldsymbol{T}_k^{\mathrm{B}}\left(\boldsymbol{K}^{\mathrm{TE}}\right)^{\mathrm{T}}\right)_w,\end{aligned}\tag{40}$$

$$\begin{aligned}&v_k\sum_{j=0}^{q}\sum_{i=0}^{n}\left(\boldsymbol{T}_k^{\mathrm{B}}\right)_{j,i}\left[\int_0^1\left[dB_{j,q}(\xi)/d\xi\right]B_{w,q}(\xi)\,d\xi\right]\\&\times\left[\int_0^1 B_{i,n}(\tau)\,B_{u,n}(\tau)\,d\tau\right]=v_k\left(\boldsymbol{K}^{\mathrm{SP}}\boldsymbol{T}_k^{\mathrm{B}}\boldsymbol{M}^{\mathrm{TE}}\right)_{w,u},\end{aligned}\tag{41}$$

$$\begin{aligned}&\alpha\sum_{j=0}^{q}\sum_{i=0}^{n}\left(\boldsymbol{T}_k^{\mathrm{B}}\right)_{j,i}\left[\int_0^1 B_{j,q}(\xi)\,B_{w,q}(\xi)\,d\xi\right]\\&\times\left[\int_0^1 B_{i,n}(\tau)B_{u,n}(\tau)\,d\tau\right]=\alpha\left(\boldsymbol{M}^{\mathrm{SP}}\boldsymbol{T}_k^{\mathrm{B}}\boldsymbol{M}^{\mathrm{TE}}\right)_{w,u},\end{aligned}\tag{42}$$

$$\alpha T_{\mathrm{amb}}\left[\int_0^1 B_{w,q}(\xi)d\xi\right]\left[\int_0^1 B_{u,n}(\tau)d\tau\right]=\alpha T_{\mathrm{amb}}\left(\boldsymbol{F}^{\mathrm{ST}}\right)_{w,u},\tag{43}$$

Combining these terms, the original PDE is transformed into the following system of ODEs for the time-dependent coefficient matrix $\boldsymbol{T}_k^{\mathrm{B}}$:

$$\begin{aligned}&1/\Delta t\cdot\boldsymbol{M}^{\mathrm{SP}}\boldsymbol{T}_k^{\mathrm{B}}\left(\boldsymbol{K}^{\mathrm{TE}}\right)^{\mathrm{T}}+v\boldsymbol{K}^{\mathrm{SP}}\boldsymbol{T}_k^{\mathrm{B}}\boldsymbol{M}^{\mathrm{TE}}\\&+\alpha\boldsymbol{M}^{\mathrm{SP}}\boldsymbol{T}_k^{\mathrm{B}}\boldsymbol{M}^{\mathrm{TE}}=\alpha T_{\mathrm{amb}}\boldsymbol{F}^{\mathrm{ST}}\ \ \forall k\in\mathcal{K},\end{aligned}\tag{44}$$

where

$$\boldsymbol{M}_{j,w}^{\mathrm{SP}}=\int_0^1 B_{j,q}(\xi)B_{w,q}(\xi)d\xi,\ \forall q\in\mathbb{N}_0,\tag{45}$$

$$\boldsymbol{M}_{i,u}^{\mathrm{TE}}=\int_0^1 B_{i,n}(\tau)B_{u,n}(\tau)d\tau,\ \forall n\in\mathbb{N}_0,\tag{46}$$

$$\boldsymbol{K}_{j,w}^{\mathrm{SP}}=\int_0^1\left[dB_{j,q}(\xi)/d\xi\right]B_{w,q}(\xi)d\xi,\ \forall q\in\mathbb{N}_0,\tag{47}$$

$$\boldsymbol{K}_{u,i}^{\mathrm{TE}}=\int_0^1\left[dB_{i,n}(\tau)/d\tau\right]B_{u,n}(\tau)d\tau,\ \forall q\in\mathbb{N}_0,\tag{48}$$

$$\boldsymbol{F}_{w,u}=\left[\int_0^1 B_{w,q}(\xi)d\xi\right]\left[\int_0^1 B_{u,n}(\tau)d\tau\right],\ \forall q,n\in\mathbb{N}_0.\tag{49}$$

The mass matrices $\boldsymbol{M}^{\mathrm{SP}}/\boldsymbol{M}^{\mathrm{TE}}$, advection matrices $\boldsymbol{K}^{\mathrm{SP}}/\boldsymbol{K}^{\mathrm{TE}}$, and load vector $\boldsymbol{F}^{\mathrm{ST}}$ are constant matrices calculated from the integrals of the Bernstein basis polynomials.

The boundary and initial conditions for the pipelines are specified as follows.

$$\boldsymbol{T}_1^{\mathrm{B}}\Big|_{(:,1)}=\boldsymbol{T}_{\mathrm{ini}}^{\mathrm{B}},\ \boldsymbol{T}_k^{\mathrm{B}}\Big|_{(1,:)}=\boldsymbol{T}_{\mathrm{in},k}^{\mathrm{B}},\ \forall k\in\mathcal{K},\tag{50}$$

where $\boldsymbol{T}_{\mathrm{ini}}^{\mathrm{B}}$ represents the initial temperature profile along the pipeline. $\boldsymbol{T}_{\mathrm{in},k}^{\mathrm{B}}$ represents the inlet temperature, which is either the supply temperature from a heat source or is determined by the energy balance among the pipes downstream of a junction.

*C. Resulting Finite-dimensional Model*

The proposed methodology converts the original infinite-dimensional variational problem into a finite-dimensional algebraic model. This final formulation is defined over the Bernstein coefficients for each time segment $k\in\mathcal{K}$.

*1) Algebraic Objective Function*

The final objective function is a quadratic function of the Bernstein coefficients.

$$\min J^{\mathrm{B}}=\sum_{i\in\mathcal{I}^{\mathrm{CU}}}C_i^{\mathrm{CU,B}}+\sum_{i\in\mathcal{I}^{\mathrm{wind}}}C_i^{\mathrm{wind,B}}+\sum_{i\in\mathcal{I}^{\mathrm{CHP}}}C_i^{\mathrm{CHP,B}},\tag{51}$$

$$C_i^{\mathrm{CU,B}}=\sum_{k\in\mathcal{K}}\left[c_{0,i}+\frac{c_{1,i}\mathbf{1}^{\mathrm{T}}\boldsymbol{P}_{i,k}^{\mathrm{B}}}{n+1}+\frac{c_{2,i}}{(n+1)^2}\left(\mathbf{1}^{\mathrm{T}}\boldsymbol{P}_{i,k}^{\mathrm{B}}\right)^2\right],\ \forall i\in\mathcal{I}^{\mathrm{CU}},\tag{52}$$

$$C_i^{\mathrm{wind,B}}=\sum_{k\in\mathcal{K}}\sigma_i\left(\bar{P}_{i,k}^{\mathrm{w}}-\frac{1}{n+1}\mathbf{1}^{\mathrm{T}}\boldsymbol{P}_{i,k}^{\mathrm{w,B}}\right),\ \forall i\in\mathcal{I}^{\mathrm{wind}},\tag{53}$$

$$\begin{aligned}C_i^{\mathrm{CHP,B}}=\sum_{k\in\mathcal{K}}\Bigg[&a_{0,i}+\frac{a_{1,i}}{n+1}\mathbf{1}^{\mathrm{T}}\boldsymbol{P}_{i,k}^{\mathrm{B}}+\frac{a_{2,i}}{n+1}\mathbf{1}^{\mathrm{T}}\boldsymbol{Q}_{i,k}^{\mathrm{B}}\\&+\frac{a_{3,i}}{(n+1)^2}\left(\mathbf{1}^{\mathrm{T}}\boldsymbol{P}_{i,k}^{\mathrm{B}}\right)^2+\frac{a_{4,i}}{(n+1)^2}\left(\mathbf{1}^{\mathrm{T}}\boldsymbol{Q}_{i,k}^{\mathrm{B}}\right)^2\Bigg],\ \forall i\in\mathcal{I}^{\mathrm{CHP}},\end{aligned}\tag{54}$$

*2) Algebraic Constraints*

The final constraints of the IHPS are algebraic constraints on the Bernstein coefficients.

$$\begin{aligned}&\sum_{l\in\mathcal{S}_i^{\mathrm{PS-}}}\left(\boldsymbol{T}_{l,k}^{\mathrm{OS,B}}\boldsymbol{m}_{l,k}^{\mathrm{S,B}}\right)+\sum_{g\in\mathcal{S}_i^{\mathrm{HS}}}\left(\boldsymbol{T}_{g,k}^{\mathrm{GS,B}}\boldsymbol{m}_{g,k}^{\mathrm{G,B}}\right)\\&=\boldsymbol{T}_{i,k}^{\mathrm{NS,B}}\left(\sum_{l\in\mathcal{S}_i^{\mathrm{PS-}}}\boldsymbol{m}_{l,k}^{\mathrm{S,B}}+\sum_{g\in\mathcal{S}_i^{\mathrm{HS}}}\boldsymbol{m}_{g,k}^{\mathrm{G,B}}\right),\\&\sum_{l\in\mathcal{S}_i^{\mathrm{PR-}}}\left(\boldsymbol{T}_{l,k}^{\mathrm{OR,B}}\boldsymbol{m}_{l,k}^{\mathrm{R,B}}\right)+\sum_{d\in\mathcal{S}_i^{\mathrm{HL}}}\left(\boldsymbol{T}_{d,k}^{\mathrm{DR,B}}\boldsymbol{m}_{d,k}^{\mathrm{D,B}}\right)\\&=\boldsymbol{T}_{i,k}^{\mathrm{NR,B}}\left(\sum_{l\in\mathcal{S}_i^{\mathrm{PR-}}}\boldsymbol{m}_{l,k}^{\mathrm{R,B}}+\sum_{d\in\mathcal{S}_i^{\mathrm{HL}}}\boldsymbol{m}_{d,k}^{\mathrm{D,B}}\right),\\&\forall i\in\mathcal{N}^{\mathrm{nd}},k\in\mathcal{K},\end{aligned}\tag{55}$$

$$\begin{aligned}&\boldsymbol{T}_{l,k}^{\mathrm{IS,B}}=\boldsymbol{T}_{i,k}^{\mathrm{NS,B}},\ \forall i\in\mathcal{N}^{\mathrm{nd}},l\in\mathcal{S}_i^{\mathrm{PS+}},k\in\mathcal{K},\\&\boldsymbol{T}_{l,k}^{\mathrm{IR,B}}=\boldsymbol{T}_{i,k}^{\mathrm{NR,B}},\ \forall i\in\mathcal{N}^{\mathrm{nd}},l\in\mathcal{S}_i^{\mathrm{PR+}},k\in\mathcal{K},\end{aligned}\tag{56}$$

$$\begin{aligned}&\sum_{l\in\mathcal{S}_i^{\mathrm{PS-}}}\boldsymbol{m}_{l,k}^{\mathrm{S,B}}+\sum_{g\in\mathcal{S}_i^{\mathrm{HS}}}\boldsymbol{m}_{g,k}^{\mathrm{G,B}}=\sum_{l\in\mathcal{S}_i^{\mathrm{PS+}}}\boldsymbol{m}_{l,k}^{\mathrm{S,B}}+\sum_{d\in\mathcal{S}_i^{\mathrm{HL}}}\boldsymbol{m}_{d,k}^{\mathrm{D,B}},\\&\sum_{l\in\mathcal{S}_i^{\mathrm{PR-}}}\boldsymbol{m}_{l,k}^{\mathrm{R,B}}+\sum_{d\in\mathcal{S}_i^{\mathrm{HL}}}\boldsymbol{m}_{d,k}^{\mathrm{D,B}}=\sum_{l\in\mathcal{S}_i^{\mathrm{PR+}}}\boldsymbol{m}_{l,k}^{\mathrm{R,B}}+\sum_{g\in\mathcal{S}_i^{\mathrm{HS}}}\boldsymbol{m}_{g,k}^{\mathrm{G,B}},\\&\forall i\in\mathcal{N}^{\mathrm{nd}},k\in\mathcal{K},\end{aligned}\tag{57}$$

$$\begin{aligned}&\boldsymbol{pr}_{i1,k}^{\mathrm{S,B}}-\boldsymbol{pr}_{i2,k}^{\mathrm{S,B}}=\mu_l\left(\boldsymbol{m}_{l,k}^{\mathrm{S,B}}\right)^2,\ \boldsymbol{pr}_{i1,k}^{\mathrm{R,B}}-\boldsymbol{pr}_{i2,k}^{\mathrm{R,B}}=\mu_l\left(\boldsymbol{m}_{l,k}^{\mathrm{R,B}}\right)^2,\\&\forall i1=\mathrm{Nd}_l^{\mathrm{in}},i2=\mathrm{Nd}_l^{\mathrm{out}},l\in\mathcal{N}^{\mathrm{pipe}},k\in\mathcal{K},\end{aligned}\tag{58}$$

$$\begin{aligned}&\boldsymbol{P}_{j,k}^{\mathrm{HP,B}}=1/\rho\eta_j\cdot\boldsymbol{m}_{j,k}^{\mathrm{G,B}}\left(\boldsymbol{pr}_{i,k}^{\mathrm{S,B}}-\boldsymbol{pr}_{i,k}^{\mathrm{R,B}}\right),\\&\forall j\in\mathcal{I}^{\mathrm{HP}},\ k\in\mathcal{K},\ i\in\mathrm{Nd}_j^{\mathrm{HS}},\end{aligned}\tag{59}$$

$$\begin{aligned}&\boldsymbol{P}_{g,k}^{\mathrm{B}}=\sum_{\zeta=1}^{NK_g}\alpha_{g,k}^{\zeta}\boldsymbol{P}_{g,k}^{\mathrm{B},\zeta},\ \ \boldsymbol{Q}_{g,k}^{\mathrm{B}}=\sum_{\zeta=1}^{NK_g}\alpha_{g,k}^{\zeta}\boldsymbol{Q}_{g,k}^{\mathrm{B},\zeta},\\&\forall g\in\mathcal{I}^{\mathrm{CHP}},\ k\in\mathcal{K},\ \sum_{\zeta=1}^{NK_g}\alpha_g^{\zeta}=1,\\&\alpha_{g,k}^{\zeta}\in[0,\ 1],\ \zeta\in\{1,2,\cdots NK_g\},\end{aligned}\tag{60}$$

$$\boldsymbol{Q}_{g,k}^{\mathrm{B}}=c\cdot\boldsymbol{m}_{g,k}^{\mathrm{G,B}}\cdot\left(\boldsymbol{T}_{g,k}^{\mathrm{GS,B}}-\boldsymbol{T}_{g,k}^{\mathrm{GR,B}}\right),\ \forall g\in\mathcal{S}^{\mathrm{HS}},k\in\mathcal{K},\tag{61}$$

$$\boldsymbol{Q}_{d,k}^{\mathrm{B}}=c\cdot\boldsymbol{m}_{d,k}^{\mathrm{D,B}}\cdot\left(\boldsymbol{T}_{d,k}^{\mathrm{DS,B}}-\boldsymbol{T}_{d,k}^{\mathrm{DR,B}}\right),\ \forall d\in\mathcal{S}^{\mathrm{HL}},\ k\in\mathcal{K},\tag{62}$$

$$\begin{aligned}&\sum_{i\in\mathcal{I}^{\mathrm{CU}}\cup\mathcal{I}^{\mathrm{CHP}}}\boldsymbol{P}_{i,k}^{\mathrm{B}}+\sum_{i\in\mathcal{I}^{\mathrm{wind}}}\boldsymbol{P}_{i,k}^{\mathrm{w,B}}-\sum_{j\in\mathcal{I}^{\mathrm{WP}}}\boldsymbol{P}_{j,k}^{\mathrm{PUMP,B}}\\&=\sum_{b\in\mathcal{N}^{\mathrm{bus}}}\boldsymbol{LD}_{b,k}^{\mathrm{B}},\ \forall k\in\mathcal{K},\end{aligned}\tag{63}$$

$$\begin{aligned}&\Bigg|\frac{1}{n+1}\sum_{b\in\mathcal{N}^{bus}}SF_{l,b}\Big(\sum_{i\in\mathcal{S}_b^{\mathrm{CU}}\cup\mathcal{S}_b^{\mathrm{CHP}}}\mathbf{1}^{\mathrm{T}}\boldsymbol{P}_{i,k}^{\mathrm{B}}+\sum_{i\in\mathcal{S}_b^{\mathrm{wind}}}\mathbf{1}^{\mathrm{T}}\boldsymbol{P}_{i,k}^{\mathrm{w,B}}-\\&\sum_{j\in\mathcal{S}_b^{\mathrm{WP}}}\mathbf{1}^{\mathrm{T}}\boldsymbol{P}_{j,k}^{\mathrm{PUMP,B}}-\sum_{b\in\mathcal{N}^{\mathrm{bus}}}\mathbf{1}^{\mathrm{T}}\boldsymbol{LD}_{b,k}^{\mathrm{B}}\Big)\Bigg|\le F_l,\\&\forall l\in\mathcal{I}^{\mathrm{tl}},\ k\in\mathcal{K},\end{aligned}\tag{64}$$

$$\underline{P}_i\le\boldsymbol{P}_{i,k}^{\mathrm{B}}\le\bar{P}_i,\ \forall i\in\mathcal{I}^{\mathrm{CU}}\cup\mathcal{I}^{\mathrm{CHP}},\ \forall k\in\mathcal{K},\tag{65}$$

$$0\le\boldsymbol{P}_{i,k}^{\mathrm{w,B}}\le\bar{P}_i^{\mathrm{w}},\ \forall i\in\mathcal{I}^{\mathrm{wind}},\ k\in\mathcal{K},\tag{66}$$

$$\begin{aligned}&0\le\boldsymbol{ru}_{i,k}^{\mathrm{B}}\le r_i^{\mathrm{up}}\cdot\Delta t,\ \forall i\in\mathcal{I}^{\mathrm{CU}},k\in\mathcal{K},\\&\boldsymbol{ru}_{i,k}^{\mathrm{B}}\le\mathbf{1}\bar{P}_i-\boldsymbol{P}_{i,k}^{\mathrm{B}},\ \forall i\in\mathcal{I}^{\mathrm{CU}},k\in\mathcal{K},\end{aligned}\tag{67}$$

$$\begin{aligned}&0\le\boldsymbol{rd}_{i,k}^{\mathrm{B}}\le r_i^{\mathrm{dn}}\cdot\Delta t,\ \forall i\in\mathcal{I}^{\mathrm{CU}},k\in\mathcal{K},\\&\boldsymbol{rd}_{i,k}^{\mathrm{B}}\le\boldsymbol{P}_{i,k}^{\mathrm{B}}-\mathbf{1}\underline{P}_i,\ \forall i\in\mathcal{I}^{\mathrm{CU}},k\in\mathcal{K},\end{aligned}\tag{68}$$

$$\begin{aligned}&\sum_{\forall i\in\mathcal{I}^{\mathrm{CU}}}\mathbf{1}^{\mathrm{T}}\boldsymbol{ru}_{i,k}^{\mathrm{B}}/(n+1)\ge SR^{\mathrm{up}},\ \forall k\in\mathcal{K},\\&\sum_{\forall i\in\mathcal{I}^{\mathrm{CU}}}\mathbf{1}^{\mathrm{T}}\boldsymbol{rd}_{i,k}^{\mathrm{B}}/(n+1)\ge SR^{\mathrm{dn}},\ \forall k\in\mathcal{K},\end{aligned}\tag{69}$$

$$-r_i^{\mathrm{dn}}\le\boldsymbol{D}_n^{\mathrm{B}}\cdot\boldsymbol{P}_{i,k}^{\mathrm{B}}\le r_i^{\mathrm{up}},\ \forall i\in\mathcal{I}^{\mathrm{CU}}\cup\mathcal{I}^{\mathrm{CHP}},k\in\mathcal{K}.\tag{70}$$

The constraints (55)-(70) constitute the equivalent algebraic representation of the non-PDE constraints of the system. Together with the Bernstein–Galerkin-based algebraic relations in (44)-(50), these constraints form the finite-dimensional representation of the original continuous spatiotemporal IHPS scheduling model. Since the VF-VT operation preserves the bilinear heat-transfer relations in the above constraints, this finite-dimensional model is formulated as a nonlinear programming problem that can be handled by solvers.

Notably, the bilinear terms are retained in their original algebraic forms. Although convex relaxation techniques, e.g., McCormick envelopes, can recast the problem into convex formulations, they may introduce relaxation errors unless the exactness of relaxation is attained. Therefore, the bilinear terms are retained to accurately describe the original nonlinear heat-transfer relations.

## V. Case Studies

In this section, the proposed BGM is numerically compared with the prevailing methods, using a single pipeline and two IHPSs of different scales. The proposed BGM is formulated in the continuous-time Bernstein function space, whereas the benchmarked methods are implemented using their original discrete-time formulations. This comparison is intended to evaluate the continuous-time Bernstein modeling paradigm against representative discrete-time dynamic DHN models. The Bernstein degree is set to $n$=3. All tests are conducted in MATLAB R2023a on a computer with 16 GB of RAM and an Intel Core i5-13400 CPU.

### *A. Simulation Conducted Using a Single Pipeline*

To validate the accuracy of thermal dynamics modeling, the BGM is compared with three prevailing approaches on a single heat pipeline, i.e., the NM [19], the discretized MOC [18], and the FDM [15]. The detailed data of the test pipeline are given in [35]. The NM, discretized MOC, and FDM are implemented in discrete-time frameworks, whereas the BGM represents the pipe-temperature field in the continuous spatiotemporal domain. Each method's performance is evaluated by comparing the outlet temperature profiles with a high-fidelity benchmark solution obtained by applying the ideal transport delay to the inlet temperature profile.

The VF-VT scenario is characterized by a fluctuating inlet temperature and flow velocity, as shown in Fig. 1. It introduces time-varying transport delays that challenge the ability of a model to capture heat transfer dynamics. The simulated outlet temperature profiles are presented in Fig. 2, with a corresponding quantitative error analysis shown in Fig. 3.

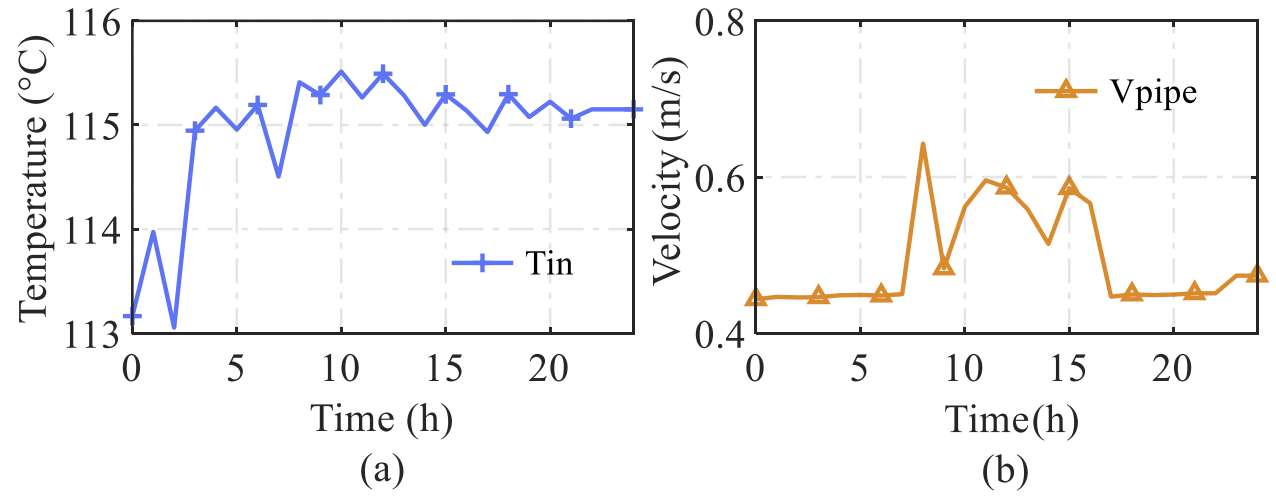

Fig. 1. Input boundary conditions for the test pipeline under VF-VT conditions: (a) inlet temperature and (b) flow velocity.

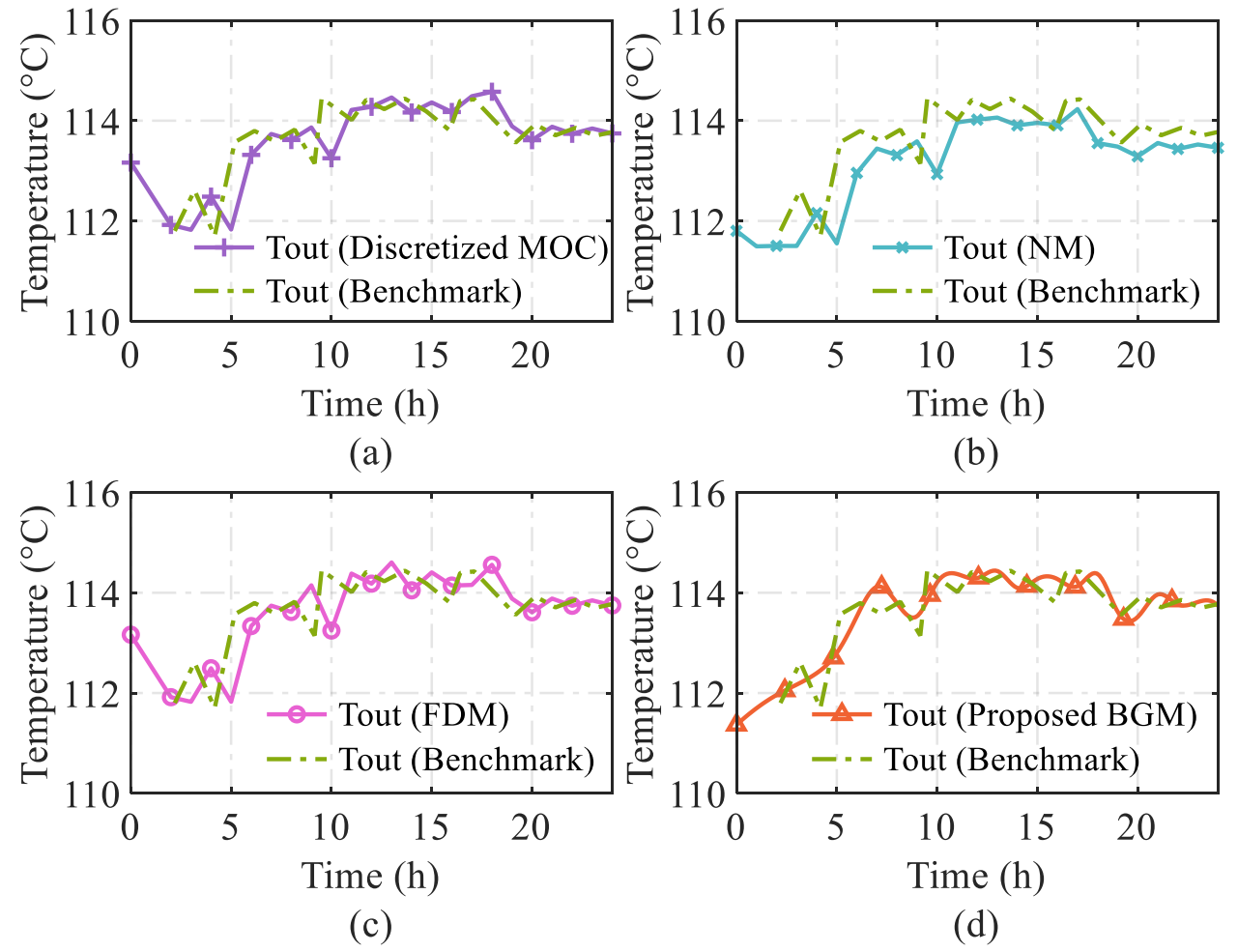

Fig. 2. Outlet temperature profiles produced for the pipeline using different methods: (a) MOC, (b) NM, (c) FDM, and (d) proposed BGM.

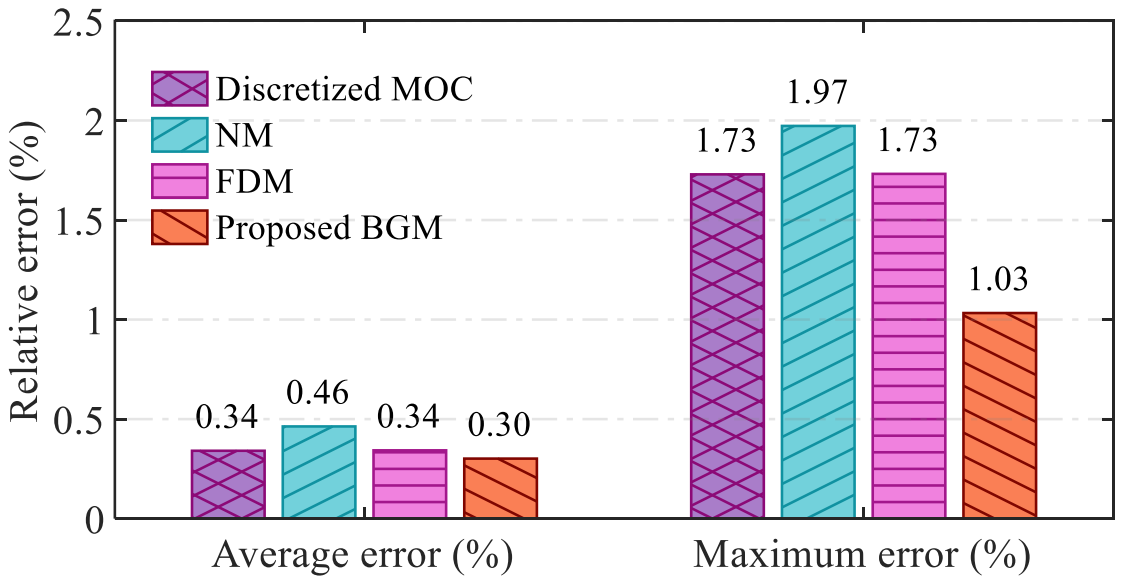

Fig. 3. Numerical error analysis of the outlet temperature: average and maximum errors induced by the discretized MOC, NM, FDM, and proposed BGM.

The differences among the methods become most pronounced during periods of rapid flow–velocity variation. As shown in Fig. 2 and Fig. 3, the fixed-grid formulations of the FDM and discretized MOC inherently suffer from numerical diffusion, leading to notable errors during these periods, whereas the NM exhibits the most severe oscillations. In contrast, the BGM closely matches the benchmark temperature profile with the smallest errors, confirming that the BGM is more accurate and robust under variable flow conditions. These results indicate that representing the pipeline temperature as a continuous spatiotemporal function helps reduce grid-induced diffusion and oscillatory artifacts when the transport velocity varies over time.

This advantage of the BGM stems from its continuous spatiotemporal formulation using Bernstein basis functions within a Galerkin framework. By projecting the PDE residuals onto the Bernstein function space, the BGM effectively reduces both numerical diffusion and oscillations, ensuring that the thermal dynamics are characterized with high fidelity.

### *B. Simulation Conducted Using a Small-Scale IHPS*

In this simulation, the proposed BGM is compared with the NM and FDM in terms of thermal modeling accuracy, operational reliability and system flexibility. All the cases use the same system settings. The proposed BGM is formulated in the continuous spatiotemporal Bernstein function space, whereas the NM and FDM are implemented in discrete-time formulation. The DHN operates under the VF-VT strategy, and the nonlinear programming models are solved via the IPOPT solver [36] with a convergence tolerance of $10^{-6}$.

The test system is an IHPS with a six-bus EPS and a six-node DHN. The configuration of this system and the associated load profiles are shown in Fig. 4. G1 is the most economical of the three units, and G2 has the highest operating cost. The CHP unit couples the EPS and the DHN, supplying both thermal and electrical energy. Detailed data are provided in [35].

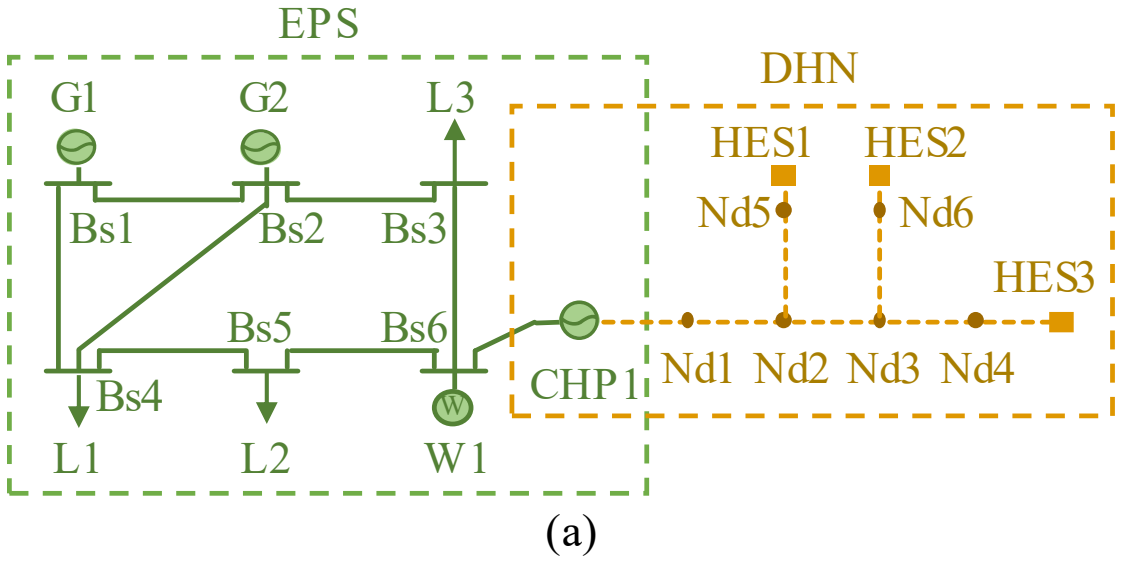

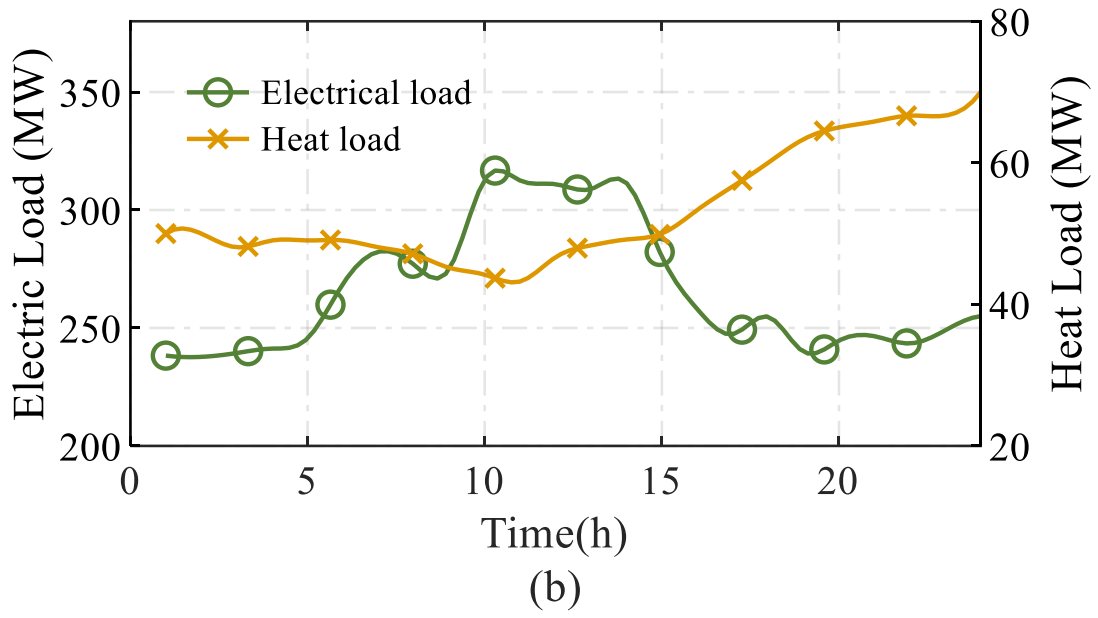

Fig. 4. (a) Configuration of the small-scale test IHPS. (b) Profiles of the electric and heat loads.

### *1) Validation of Thermal Dynamics Accuracy*

The accuracy of thermal dynamics modeling is evaluated by comparing the temperature profiles obtained via each method. As shown in Fig. 5, the BGM yields smooth temperature profiles without frequent variations, accurately capturing the thermal dynamics of the DHN. In contrast, the FDM produces noticeable numerical oscillations, particularly in the Nd4 supply temperature profile during 6–10 h. The NM performs the worst, exhibiting erratic temperature oscillations in the Nd4 return temperature profile during 11–13 h, underscoring its inability to accurately model thermal dynamics.

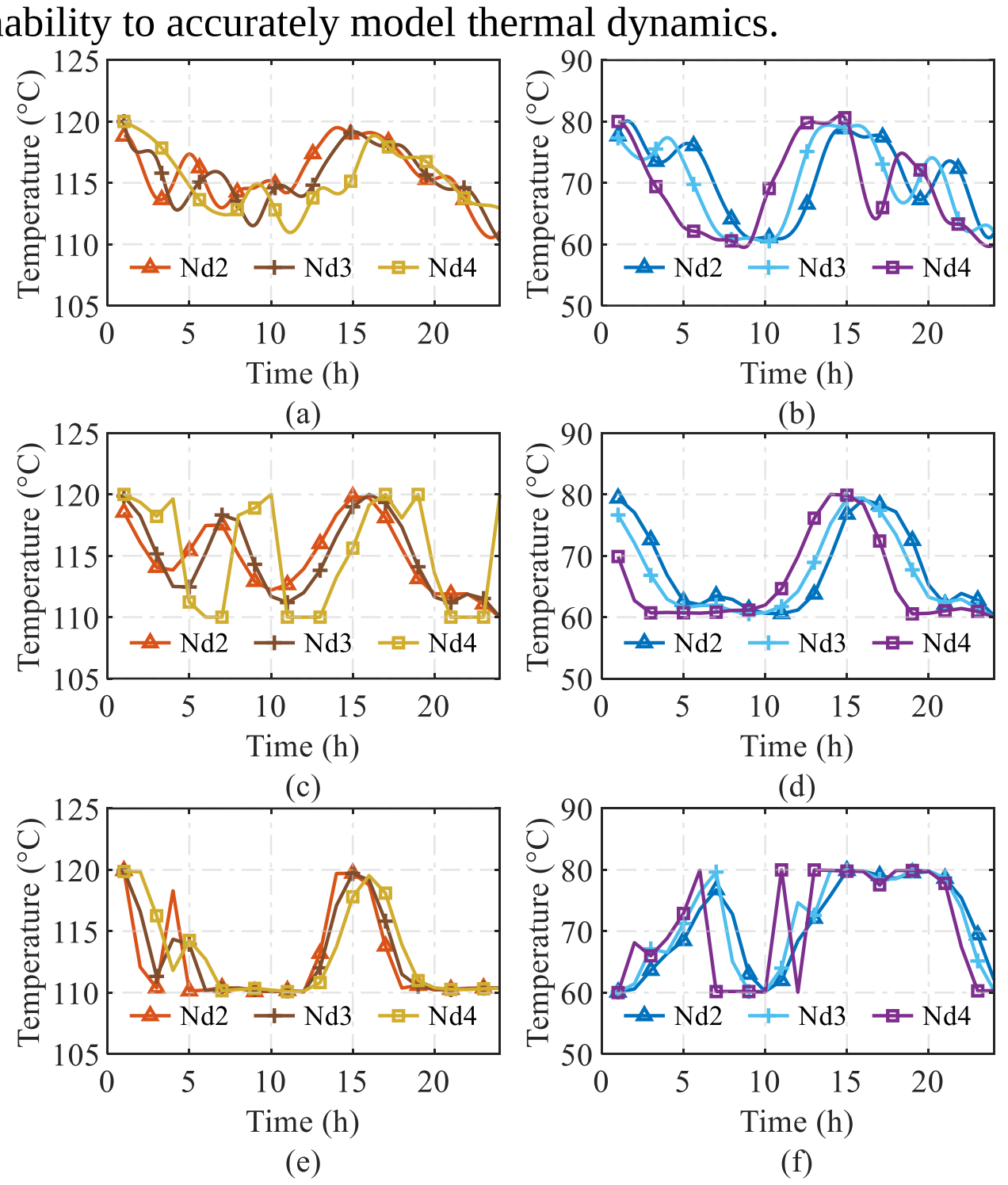

Fig. 5. Comparison among the node temperatures in the small-scale IHPS. Supply temperature: (a) proposed BGM, (c) FDM, and (e) NM. Return temperature: (b) proposed BGM, (d) FDM, and (f) NM.

To further investigate the physical fidelity of the proposed BGM, Fig. 6 shows the spatiotemporal evolution of the heat transport process across two adjacent pipeline segments, namely, Nd2–Nd3 and Nd3–Nd4. The diagonal trajectories of the thermal fronts in Figs. 6(b) and (d) present obvious transport delays; e.g., a thermal front initiated at $t$=5 h at the inlet reaches the outlet at approximately $t$=7 h. The smooth propagation of these fronts, shown in Figs. 6(a) and (c), reflects the substantial thermal inertia of the DHN, which dampens abrupt temperature fluctuations. Simultaneously, the fading color intensity on the surfaces visualizes the spatial attenuation of the peak temperatures due to heat dissipation. This process highlights network coupling, as the outlet temperature of segments Nd2–Nd3 serves as the inlet boundary condition for Nd3–Nd4, ensuring the continued propagation of thermal dynamics throughout the network.

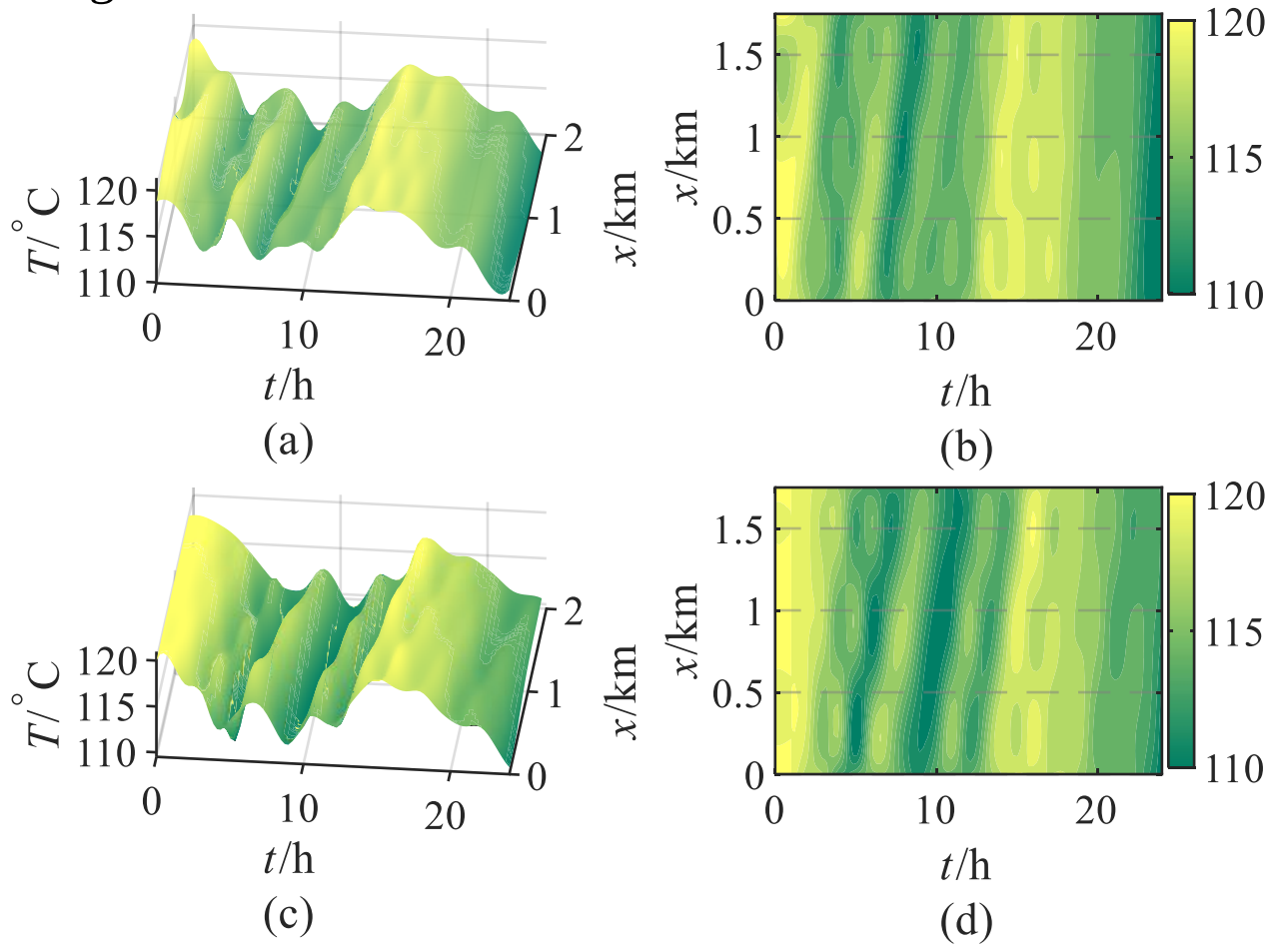

Fig. 6. Spatiotemporal pipeline temperature distribution obtained from the proposed BGM. (a), (b) Pipeline Nd2–Nd3. (c), (d) Pipeline Nd3–Nd4.

### *2) Verification Simulation of Operational Reliability*

To verify the practical operational reliability of the BGM, the dynamics of the DHN are simulated via the ODE solver embedded in MATLAB to obtain the real node temperatures across the network, with the CHP thermal output and supply temperature profiles obtained via optimization serving as time-varying boundary conditions. As shown by the low return temperature at Nd5 in Fig. 7, the simulation results indicate that the NM and the FDM lead to violations of the return temperature requirements.

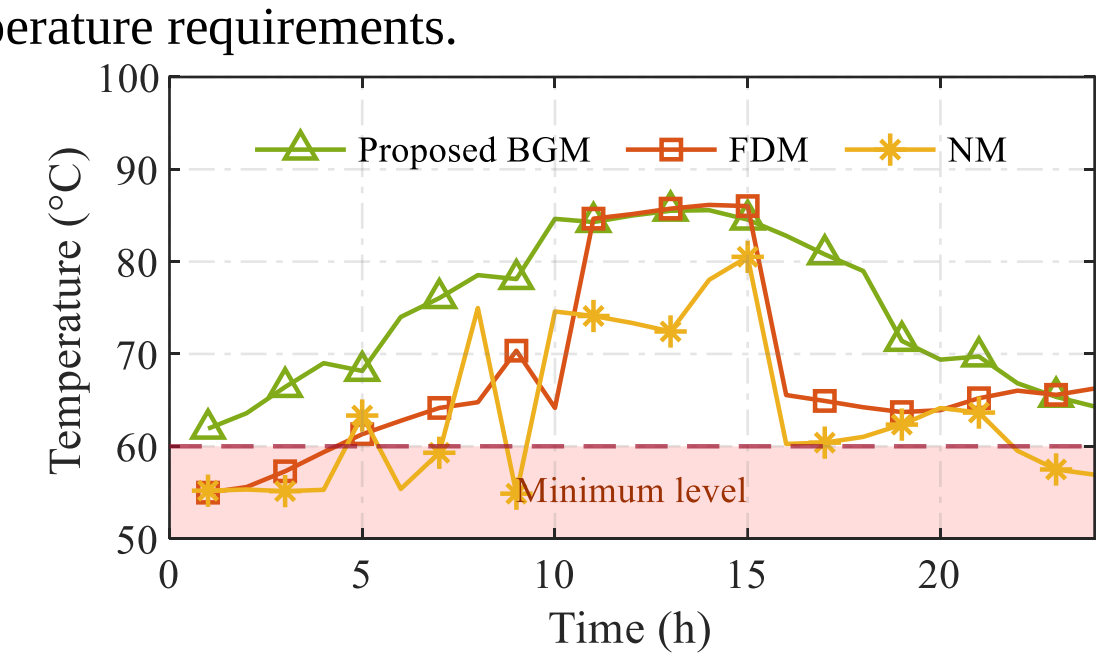

Fig. 7. Return temperature of Nd5 obtained from the simulation verification.

The operational reliability of each method is evaluated in terms of the minimum temperature constraints via the maximum underdeviation $\Delta T^{\max}_{\text{under}}$ and undersupply rate $R^{\%}_{\text{under}}$.

$$\Delta T^{\max}_{\text{under}} = \max_{i\in\mathcal{N}^{\text{nd}},k\in\mathcal{T}} \left\{ \max\left(0,\ \underline{T}_i - T_{i,k}\right)\right\}, \tag{71}$$

$$R^{\%}_{\text{under}} = 1/N_{\text{total}} \cdot \sum\nolimits_{i\in\mathcal{N}^{\text{nd}}} \sum\nolimits_{k\in\mathcal{T}} I\left(T_{i,k} < \underline{T}_i\right) \times 100\%, \tag{72}$$

where $T_{i,k}$ is the simulated temperature at node $i$, $\underline{T}_i$ denotes the lower limit, $I(\cdot)$ is an indicator function that equals 1 if its argument is true, and $N_{\text{total}}$ is the total number of spatiotemporal evaluation points. In the following analysis, $(\cdot)^{\text{S}}$ and $(\cdot)^{\text{R}}$ denote the metrics for the supply and return networks, respectively.

The results, summarized in Table I, show that the BGM leads to minor violations, whereas the FDM and NM schedules result in notable violations in the return network. Specifically, the NM schedule results in a maximum temperature decrease of 8.707°C and an undersupply rate of 27.778%, highlighting the superior reliability of the BGM for practical applications. This finding indicates that thermal–dynamic errors introduced by discrete-time approximations may be transferred to the scheduling decision and lead to an infeasible or unreliable heat supply. The continuous space–time representation reduces this mismatch by preserving the continuous propagation of pipeline temperatures within the scheduling model.

TABLE I
THERMAL RELIABILITY COMPARISON FOR THE SMALL-SCALE IHPS

| Method | $\Delta T_{\max}^{\text{under, S}}$ (°C) | $R_{\%}^{\text{under, S}}$ (%) | $\Delta T_{\max}^{\text{under, R}}$ (°C) | $R_{\%}^{\text{under, R}}$ (%) |
|---|---|---|---|---|
| BGM | 0 | 0 | 0.625 | 1.389 |
| FDM | 0 | 0 | 8.529 | 9.028 |
| NM | 0.052 | 9.722 | 8.707 | 27.778 |

*3) Economic and System Flexibility Benefits Analysis*

Fundamentally, the modeling accuracy of thermal dynamics directly affects the performance of dispatch results. Compared with the NM's operation cost of $27,823.12, the BGM and FDM achieve reduced costs of $26,245.59 and $26,692.45, respectively. In terms of computational efficiency, the BGM requires 7.707 s, which is shorter than the NM and FDM runtimes of 27.94 s and 78.94 s, respectively.

This difference in cost can be attributed to the distinct dispatch strategies of each method. As shown in Fig. 8, the BGM utilizes a lower CHP output, which in turn increases the dispatch of the more economical unit G1 and reduces that of the costly unit G2. This dispatch in the BGM not only reduces the operation costs but also increases the availability of the cumulative reserve of the system to 607.8 MW, compared with 576.6 MW for the FDM and 561.8 MW for the NM. This enhanced flexibility allows the system to better accommodate uncertainties such as load fluctuations and forecasting errors, ensuring more robust and reliable operations.

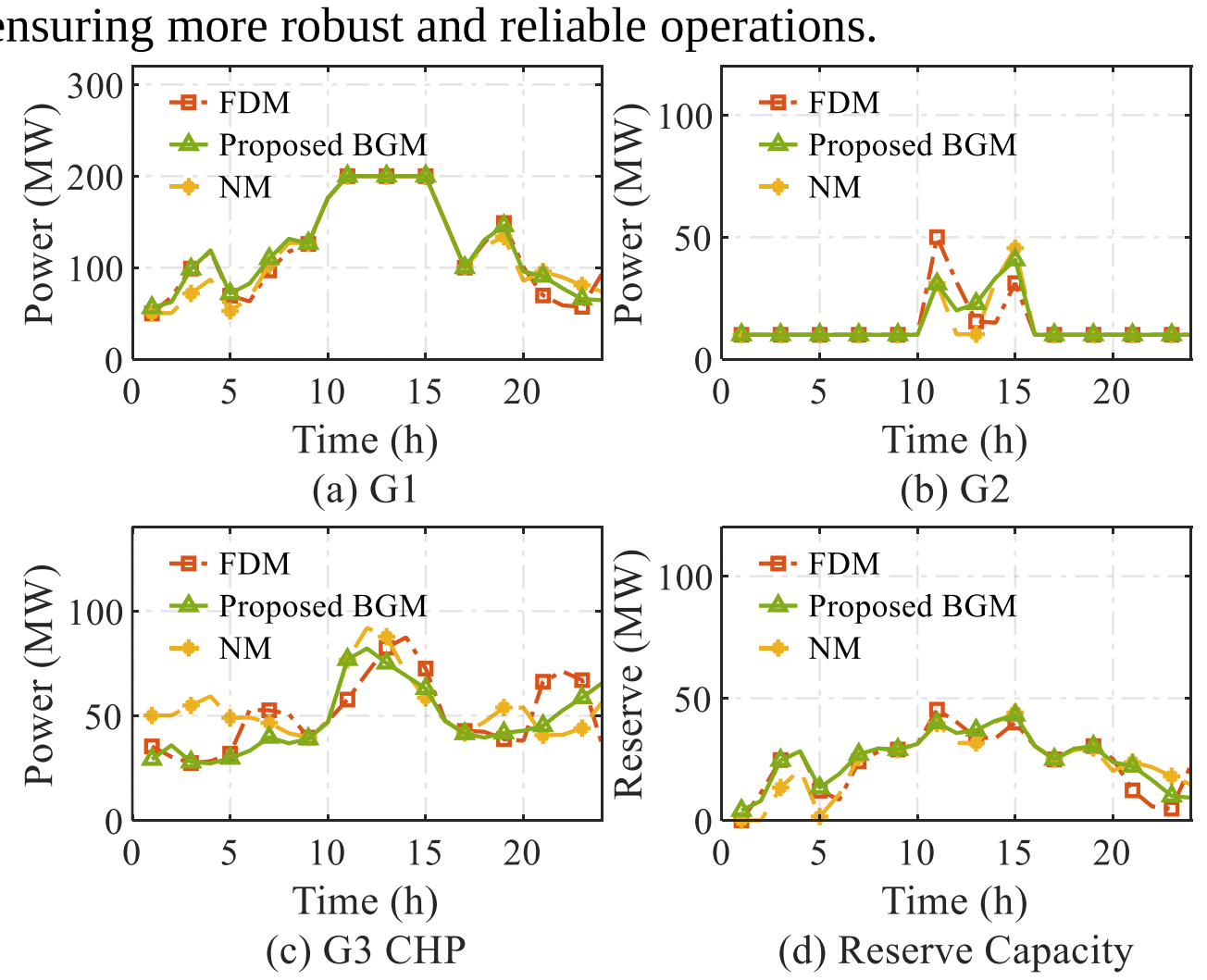

Fig. 8. Dispatch scheme comparison for the small-scale IHPS: (a) G1, (b) G2, (c) G3 (CHP), and (d) available reserve capacity.

Furthermore, unreliably modeling thermal dynamics compromises system-wide operations such as renewable energy accommodation. As shown in Fig. 9, the inaccurate dynamic representation of the NM results in a wind curtailment of 19.4 MWh. In contrast, the higher-fidelity BGM and FDM enable more accurate scheduling of thermal dynamics, thereby fully utilizing wind power generation.

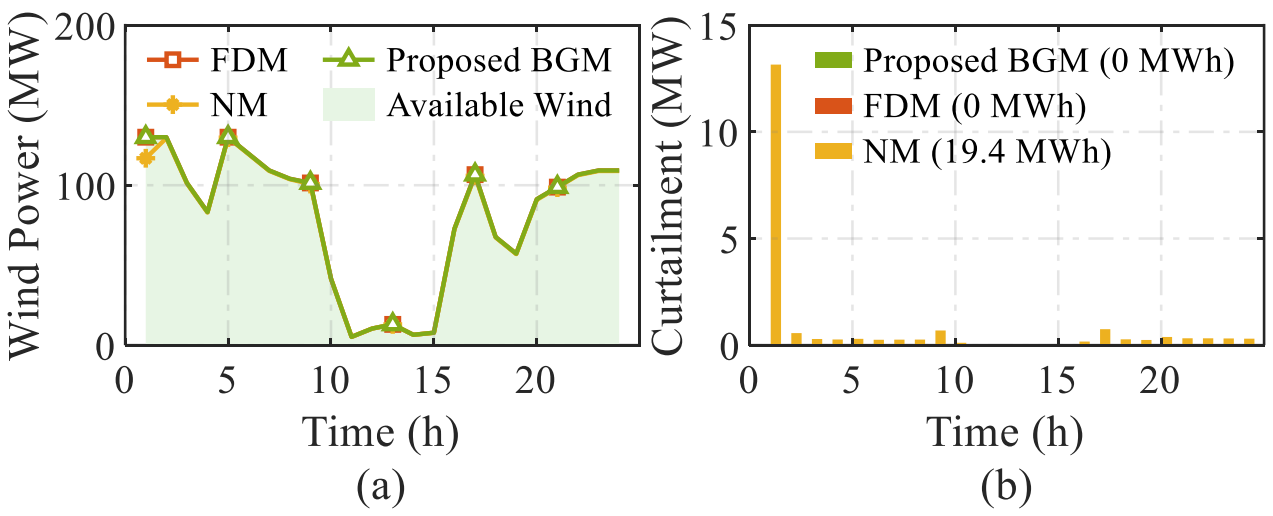

Fig. 9. Comparison between the wind outputs and curtailments in the small-scale IHPS.

*4) Impact of VF-VT Regulation*

A comparative study is conducted on the same small-scale IHPS under the CF-VT and VF-VT modes. Both scenarios share the same system settings. In the CF-VT case, the mass flow profile is fixed, and the resulting linear programming model is solved via Gurobi [37]. In the VF-VT case, both the mass flow and temperature trajectories are optimized in the scheduling model with nonlinear thermal–hydraulic coupling.

TABLE II
COMPARISON OF THE OPERATION PERFORMANCE UNDER CF–VT AND VF–VT

| Mode | Wind Curtailment (MWh) | Available Reserve Capacity (MW) | Total Cost ($) | Runtime (s) |
|---|---|---|---|---|
| CF-VT | 3.113 | 0 | 27,595.75 | 0.14 |
| VF-VT | 0 | 607.8 | 26,245.59 | 7.707 |

As shown in Table II, compared with the CF-VT operation mode, the VF-VT operation mode improves the scheduling performance. Specifically, the CF-VT operation mode results in 3.113 MWh wind curtailment and no available reserve capacity, whereas the VF-VT operation mode eliminates wind curtailment and provides 607.8 MW of available reserve capacity. Moreover, the total operating cost is reduced from $27,595.75 to $26,245.59.

The operating profiles of the CHP unit in Fig. 10 further illustrate the dispatch effect of VF-VT. Compared with the CF-VT mode, the VF-VT operation mode reduces the CHP electricity generated from 1252.9 MWh to 1146.8 MWh while increasing the average CHP supply temperature from 114.4°C to 116.2°C. This finding indicates that VF-VT satisfies the thermal demand through coordinated regulation of mass flow and temperature rather than by relying on higher CHP electric output. As a result, the heat–electricity coupling restriction on CHP dispatch is alleviated, resulting in more available reserve capacity for the EPS and avoiding wind curtailment. Although the solution time increases from 0.14 s to 7.707 s because VF-VT retains nonlinear thermohydraulic coupling, the additional computational effort remains acceptable for the obtained flexibility and cost benefits.

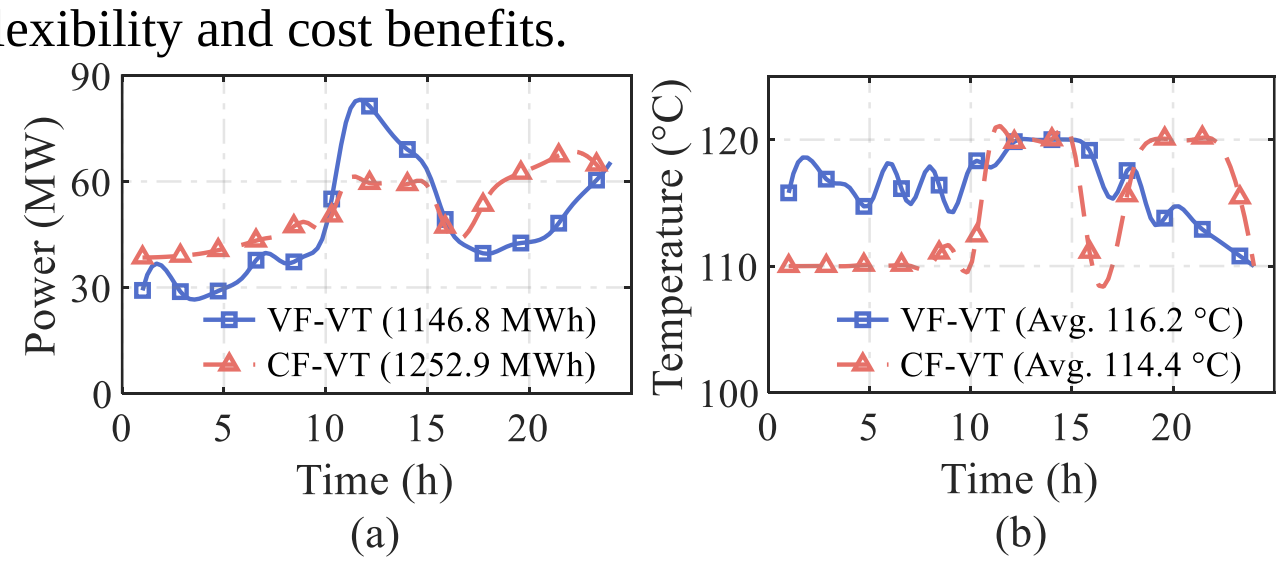

Fig. 10. Comparison between the VF-VT and CF-VT modes in the small-scale IHPS: (a) CHP power output and (b) CHP supply temperature.

### C. *Simulation Conducted Using a Large-Scale IHPS*

This test is conducted on an IHPS with a modified IEEE 39-bus EPS and a 20-node DHN. The detailed parameters are given in [35]. All tests are conducted under the VF-VT strategy and solved via IPOPT. The operating costs of the BGM, FDM and NM are $1,769,369.64, $1,818,977.99 and $1,758,588.07, respectively. The computational times required by the BGM, FDM, and NM are 268.53 s, 495.54 s, and 753.83 s, respectively. Hence, compared with the other methods, the BGM achieves a competitive operating cost while demonstrating superior computational efficiency. These results indicate that continuous spatiotemporal modeling does not inherently increase the computational burden of system-level scheduling. By parameterizing the temperature field via the Bernstein coefficients, the BGM avoids the dense grid-based computation required by the high-resolution FDM, yielding a compact formulation that drives its computational advantage in large-scale cases. In contrast, the FDM incurs a greater computational burden because of its discrete finite difference. The NM, on the other hand, is limited by insufficient physical fidelity in representing thermal dynamics and slow convergence.

TABLE III
THERMAL RELIABILITY COMPARISON FOR THE LARGE-SCALE IHPS

| Method | $\Delta T_{\max}^{\text{under, S}}$ (°C) | $R_{\%}^{\text{under, S}}$ (%) | $\Delta T_{\max}^{\text{under, R}}$ (°C) | $R_{\%}^{\text{under, R}}$ (%) |
|---|---|---|---|---|
| BGM | 0.06 | 2.5 | 0.068 | 5 |
| FDM | 0 | 0 | 9.039 | 6.25 |
| NM | 0 | 0 | 7.613 | 21.042 |

As shown in Table III, the BGM achieves an undersupply rate of 5.0%, with a maximum underdeviation temperature of 0.068°C. In contrast, the FDM and NM have substantial deficiencies, with maximum negative temperature deviations of 9.039°C and 7.613°C and considerable undersupply rates of 6.25% and 21.042%, respectively. These deficiencies severely compromise the operational reliability of the IHPS.

The wind power utilization results, illustrated in Fig. 11, further confirm that the fidelity of the thermal model is crucial for ensuring the flexibility of the release system.

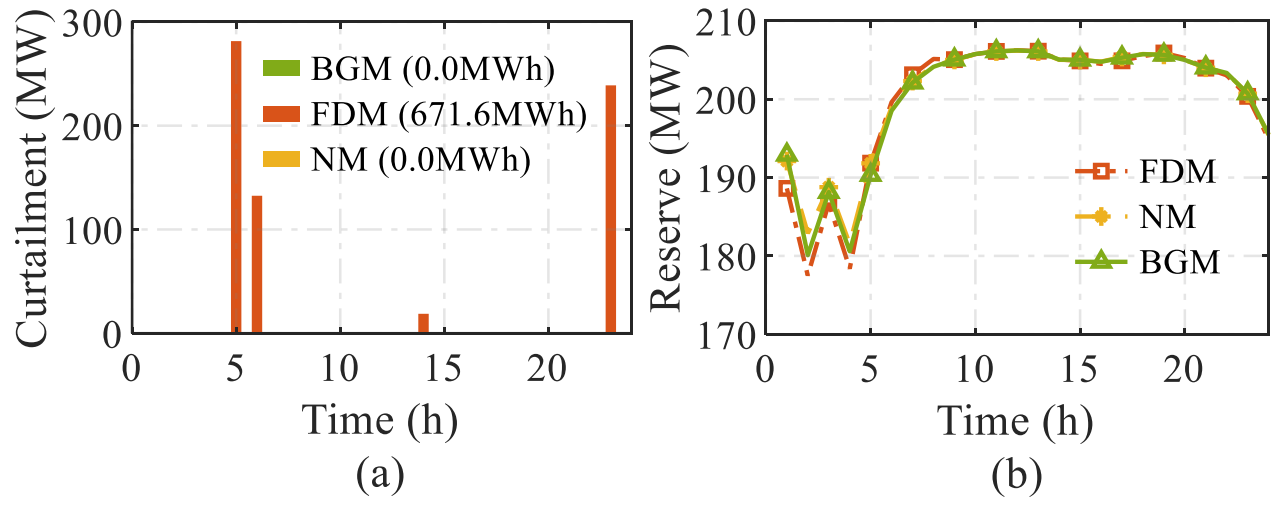

Fig. 11. Wind power curtailment and system reserve capacity comparisons for the large-scale IHPS: (a) wind power curtailment, (b) available reserve capacity of the system.

As shown in Fig. 11(a), the FDM results in severe wind curtailment values of up to 671.6 MWh, whereas the NM avoids curtailment at the cost of thermal reliability, as reflected in the previously high undersupply rates. In contrast, the BGM accommodates all available wind power while ensuring thermal reliability. Fig. 11(b) shows the available reserve capacity of the system. The FDM maintains the lowest reserve capacity of 4798.7MW because of its wind curtailment, whereas the NM and BGM maintain comparable levels t 4812.0 MW and 4806.4 MW, respectively. This higher reserve capacity of the BGM enhances the resilience of the system, safeguarding the grid against unforeseen contingencies and ensuring greater operational security.

## VI. CONCLUSION

An IHPS scheduling model with continuous-time thermal dynamics is presented in this paper, together with a Bernstein–Galerkin method via the Bernstein function space. The key advantage of the proposed approach lies in its ability to capture continuous temperature dynamics over both time and space. A single-pipe simulation demonstrates that the proposed method effectively avoids the numerical dissipation that is common in conventional discrete approaches, thereby preserving the physical thermal inertia of the DHN. Case studies involving two IHPSs at different scales demonstrate that by preserving physical fidelity, the proposed method effectively unlocks the flexibility potential of DHNs and provides enhanced economic performance.

Future work will focus on addressing the multi-time-scale characteristics that are inherent in complex energy systems. Another important challenge is the incorporation of robust safety constraints, as the time delays observed in subsystem interactions significantly affect system security. These issues remain open directions for further research.